\documentclass{emulateapj}

\usepackage{graphicx}
\usepackage{epstopdf}
\shorttitle{Gamma-ray Line Search}
\shortauthors{Teegarden et al.}

\begin{document}

\title{A Comprehensive Search for Gamma-Ray Lines in the First Year of Data from the INTEGRAL Spectrometer}

\author{B. J. Teegarden\altaffilmark{1} and K. Watanabe\altaffilmark{1,2}}  

\altaffiltext{1}{Exploration of the Universe Division, NASA/Goddard Space
Flight Center, Greenbelt, MD 20771, USA}
\altaffiltext{2}{University of Maryland, College Park, MD, USA}
%%\email{bonnard@lheamail.gsfc.nasa.gov}

\begin{abstract}
 We have carried out an extensive search for gamma-ray lines in the first year of public data from the Spectrometer (SPI) on the INTEGRAL mission.  INTEGRAL has spent a large fraction of its observing time in the Galactic Plane with particular concentration in the Galactic Center (GC) region ($\sim3$  Msec in the first year).  Hence the most sensitive search regions are in the Galactic Plane and Center.   The  phase space of the search spans the energy range 20-8000 keV and line widths from 0-1000 keV (FWHM). It includes both diffuse and point-like emission.  We have searched for variable emission on time scales down to $\sim 1000$ sec.  Diffuse emission has been searched for on a range of different spatial scales from $\sim20\degr$ (the approximate field-of-view of the spectrometer) up to the entire Galactic Plane.  Our search procedures were verified by the recovery of the known gamma-ray lines at 511 keV and 1809 keV at the appropriate intensities and significances.  We find no evidence for any previously unknown gamma-ray lines.  The upper limits range from a few  $\times10^{-5}  \mbox{ cm}^{-2} \mbox{ s}^{-1} $ to a few $\times10^{-2}  \mbox{ cm}^{-2} \mbox{ s}^{-1} $ depending on line width, energy and exposure; regions of strong instrumental background lines were excluded from the search.  Comparison is made between our results and various prior predictions of astrophysical lines. 
\end{abstract}


\section{Introduction}
\label{intro}

\subsection {Astrophysical Gamma-Ray Line Production}
\label{grl_prod}
Gamma-ray lines can be the bearers of important information on astrophysical sites and processes.  A large body of work over the past 3 decades has identified a wide variety of different astrophysical scenarios in which gamma-ray line production can occur (for recent reviews see \citet{kno00, pra04, vin05}).  The most important of these are summarized as follows:

\it Nucleosynthesis in Massive Stars: \rm   Radioactive elements are created through hydrostatic nucleosynthesis in the cores of massive stars. Long-lived isotopes (e.g. $^{26}$Al, $t_{1/2}  =7.2 \times 10^{5} $ yr, $E_{\gamma} =$1809 keV) can be convected to the surface and carried into the interstellar medium (ISM) by their stellar winds where they subsequently decay and produce gamma-ray lines.  Such lines can be valuable tracers of recent nucleosynthesis in our galaxy.

\it Supernovae (SN): \rm Radioactive nuclei are created by explosive nucleosynthesis in both Type I and Type II supernovae.  In both cases the dominant element synthesized is  $^{56}$Ni which decays to $^{56}$Co ($t_{1/2} = 6.1$ d, $E_{\gamma} =$158, 270, 480, 750, 812 keV) and then to $^{56}$Fe ($t_{1/2} = 77$ d, $E_{\gamma} =$ 847, 1238, 2599 keV). Type Ia SN are believed to be produced by either the accretion-induced collapse of a white dwarf (WD) or the merger of a WD-WD binary.  At early times the SN shell is optically thick to gamma-rays, becoming transparent after $\sim60$ days.  The strongest lines are therefore from the longer-lived decay of $^{56}$Co.  Later in the expansion even longer-lived isotopes take over (e.g. $^{57}$Co, $t_{1/2} = 272$ d, $E_{\gamma} =$122, 136 keV).  Type II SN result from the core collapse of a more massive star which has exhausted its nuclear fuel.  The envelope takes longer to become transparent and consequently the gamma-ray line strengths are generally weaker.   As with Type Ia SN the  $^{56}$Ni  $\rightarrow ^{56}$Co $\rightarrow ^{56}$Fe decay chain also dominates, followed by $^{57}$Co decay.  The former decay is believed to be the engine that powers the early-time expansion of the SN shell.   $ ^{56}$Co undergoes $\beta^{+}$-decay 19\% of the time.  Most of the positrons will annihilate in the expanding SN shell before they can escape \citep{cha93, mil99}.  However, for Type Ia SN, if the magnetic field has a sufficiently smooth and open geometry, a significant number of positrons can escape.  It is possible that these positrons can account for the observed bulge component of the Galactic 511 keV emission  \citep{pur97, kno05}.

\it Gamma-Ray Bursts/Hypernovae: \rm Hypernovae are asymmetric explosions of massive stars.  They are leading candidates for the origin of long duration gamma-ray bursts.   It has been suggested that positron production by hypernovae can be a significant contributor to the Galactic 511 keV flux \citep{cas04a,ber04}.  Positrons can be efficiently transported outward by the strong jets that are believed to accompany hypernovae  and escape into into ISM where they subsequently annihilate.

\it Novae: \rm A classical nova is believed to be caused by a thermonuclear detonation on the surface of white dwarf resulting from the build-up of material accreted from a companion star.  Unstable proton-rich isotopes such as $^{13}$N, $^{14}$O, $^{15}$O, $^{17}$F and $^{18}$F are created, which decay primarily by positron emission (see e.g. \citet{lei87,wei90,nof91}.)  Since most of these have short half-lives the positrons are created during the early phase of the expansion where densities are high, and the annihilation lifetime is short.   Although bright, the 511 keV signal from a classical nova will typically last $<$ 1 day \citep{gom98}.  The probability of INTEGRAL with its $\sim 25 \degr$ field-of-view catching such an event is small.  There are two long-lived isotopes, $^{7}$Be ($t_{1/2} = 53.1$~d, $E_{\gamma} =$ 478~keV) and $^{22}$Na ($t_{1/2} = 2.6$~y, $E_{\gamma} =$~1275~keV), which produce gamma-ray lines that might have a greater chance of being detected by INTEGRAL.  The former is primarily produced by CO novae and the latter by ONe novae.  We will return later to the issue of the detection of gamma-ray lines from novae.

\it Black Holes: \rm Electron-positron jets are believed to be characteristic features of both super-massive and stellar-mass black holes (BH) (see, for example, \citet{ree84, mir99}).  Positrons can be transported by the jets far from the BH where they eventually will annihilate in the ISM.  The 511 keV bulge component could conceivably originate from the massive BH at the center of our Galaxy or alternatively from an ensemble of stellar-mass BH's each producing positrons that migrate into the ISM and produce the diffuse emission that is observed.  In the latter scenario, however, one might expect a more pronounced disk component with a larger disk-to-bulge ratio than is observed \citep{kno05}.

\it Pulsars: \rm Electron-positron pairs are believed to be created by vacuum pair production in the strong magnetic fields in pulsar magnetospheres (see, for example, \citet{rud72}).  MeV positrons that escape the magnetosphere will take typically  a few million years to annihilate.  Depending on the geometry and smoothness of the local Galactic magnetic field the positrons could migrate far from the source before they annihilate.  The resultant 511 keV emission could manifest itself as either an ensemble of point-like sources or  a diffuse glow.   

\it Neutron Star Binaries: \rm Matter accreting onto a neutron star surface can produce gamma-ray lines by a  variety of interactions.  Probably the strongest of these is neutron-capture on hydrogen which produces a line at 2223 keV \citep{bil93}. 

\it Cosmic Ray Interactions: \rm Cosmic rays can activate nuclei in the ISM which subsequently decay and produce gamma-ray lines.  Positrons can also be created through either  $\beta^{+}$-decay or $\pi^+$ production($\pi^{+ }\rightarrow \mu^{+} \rightarrow e^{+})$.  The most extensive treatment of cosmic-ray produced gamma-ray lines can be found in \citet{ram79}.  The authors have exhaustively inventoried the great number of different possible interactions and decay channels.  The predicted gamma-ray spectra display a rich variety of lines in the energy range from a few hundred keV to 8 MeV.  Narrow lines are normally produced by the interaction of cosmic-ray ray protons with ambient interstellar nuclei.  Broadened lines are produced when cosmic-ray heavy nuclei collide with ambient interstellar hydrogen.   Narrow lines are easier to detect with a high-resolution instrument such as INTEGRAL/SPI.  The strongest narrow lines are those from electron-positron annihilation (511 keV), $^{56}$Fe excitation (847 keV),  $^{12}$C excitation (4439 keV) and $^{16}$O excitation (6129 keV).  Unfortunately these line strengths are near the limits of detectability with current instrumentation.   We will return to this point later with a comparison of our limits and the predictions of theory.

\it Dark Matter: \rm There has been recent speculation on the possible existence of light ($m <~100 \mbox{~MeV} $) dark matter particles which would decay or annihilate primarily through the formation of electrons and positrons \citep{boe04, cas04b, hoo04, oak04, pic04,scha04}.  A variety of different particles including axions, sterile neutrinos and quark droplets have been proposed.  Dark matter decay or annihilation could possibly account for the bulge component of the electron-positron annihilation radiation.  However, \citet{bea04} have pointed out that dark matter annihilation or decay will produce continuum gamma-rays as well as 511 keV.  The former would exceed the observed gamma-ray flux from the Galactic Center for all but the lightest dark matter particles.  

\subsection{Known Astrophysical Gamma-Ray Lines}
\label{grl_known}

Table~\ref{obs_lines}  summarizes the known astrophysical gamma-ray lines. The number of actually observed lines is small compared to the number of predictions.   This may reflect an excess of optimism on the part of the theoreticians or insufficient sensitivity on the part of previous instruments.  INTEGRAL/SPI represents a significant step forward in its ability to detect and study gamma-ray lines.  Hence there is considerable motivation to thoroughly search it's data for new lines, which is the subject of this paper.  Search variables include line energy, width, temporal and spatial distribution.  INTEGRAL/SPI combines fine energy resolution with moderate spatial resolution.  This, along with the need to examine different time scales (since most of the point sources are variable), leads to a very large volume of phase space to be searched.  We elaborate the properties of the known lines as follows.

\subsubsection{Electron-Positron Annihilation}
\label{ep_ann}

The 511 keV electron-positron annihilation line from the GC was the first-discovered and remains the strongest.   First detected in early balloon flights \citep{joh72,lev78}, it has subsequently been studied in detail by several satellites [HEAO-C \citep{rie81}, WIND/TGRS \citep{tee96}, CGRO/OSSE\citep{pur97}].  INTEGRAL \citep{jea04,tee05,kno05} has given us the most recent and most detailed information.  The results are all generally consistent with a narrow ($\sim2$ keV FWHM) unshifted line with the bulk of the emission coming from a $\sim10\degr$ diameter region located at the Galactic Center.  However, \citet{jea05} have recently reported evidence for a second broadened component of the 511 keV line such as would be expected from annihilation of positronium formed in-flight.  The combined INTEGRAL measurements of the 511 keV line profile and associated positronium continuum imply that the annihilation took place in the warm partially-ionized component of the interstellar medium \citep{chu05,jea05}.  OSSE and INTEGRAL also marginally detect a more extended Galactic Plane component \citep{pur97, kno05}.  In an early analysis OSSE \citep{pur97} reported an enhancement of 511 keV emission above the Galactic Plane (the so-called "annihilation fountain").  However, later  OSSE analyses \citep{kin01,mil01} found little or no evidence for this component, and INTEGRAL does not detect it \citep{jea04,kno05}.  The strong concentration of emission and large bulge-to-disk ratio imply that the Galactic 511 keV emission comes from an older Galactic population, for example SNIa or LMXBs \citep{kno05}.

In 1990-91 the SIGMA experiment on the Granat satellite found evidence for transient line-like features in the spectra of two different sources 1E1740.7-2942 \citep{bou91} and Nova Musca \citep{gol92}.  In 1E1740.7-2942 a strong (flux~$= 1.2 \times 10^{-2} \mbox{ cm}^{-2} \mbox{ s}^{-1} $) broadened (FWHM = 240 keV) feature was found at an energy of $480 \pm 90$ keV lasting for $\sim$ 2 days.  It has been interpreted as (potentially) red-shifted $e^{+}$ - $e^{-}$ annihilation radiation.  Four months later a similar event was observed from Nova Musca.  The feature lasted $\sim$ half a day.  The measured line parameters were, flux~$= 6 \pm 3 \times  10^{-3} \mbox{ cm}^{-2} \mbox{ s}^{-1} $, E = $481 \pm 22$ keV, and FWHM~$= 23 \pm 23$~keV.  Other wide-field instruments with comparable or better sensitivities (SMM, BATSE), however, have not found such transients \citep{sha93,che98,smi96}.  There was also a simultaneous observation of Nova Musca by OSSE during one of the reported Granat/SIGMA transients in which no transient lines were seen \citep{jun95}

\subsubsection{$^{26}Al$}
\label{26al}

The 1809 keV line from radioactive $^{26}$Al was first detected by HEAO-C \citep{mah84}.  The COMPTEL experiment on CGRO produced detailed maps showing a distribution narrowly confined to the Galactic Plane with local hot spots in the Cygnus and Vela regions.  The distribution was much more extended in the Galactic Plane than the 511-keV emission and was generally consistent with that expected for young massive stars \citep{die95,kno99}.   A broadened line ($6.4 \pm 1.2$ keV FWHM) was reported by the GRIS experiment \citep{nay96}, however subsequent observations by RHESSI \citep{smi04} and INTEGRAL \citep{die03b}  found no evidence for any line broadening. $^{26}$Al undergoes  $\beta^{+}$-decay 82\% of the time.  The resultant 511 keV emission from the annihilation of these positrons is $\sim50\%$ of the observed 511 keV disk component \citep{kno05}, but within the (large) errors consistent with it.

\subsubsection{$^{44}Ti$} 
\label{44ti}

$^{44}$Ti is synthesized in core-collapse supernovae.  It decays via the chain $^{44}\mbox{Ti}$  $\rightarrow$  $^{44}\mbox{Sc}$ $ \rightarrow$   $^{44}\mbox{Ca}$ with a half-life of 63 yr and in the process produces gamma-ray lines at 68, 78 and 1157 keV, with nearly the same intensity.    $^{44}$Ti lines then are signatures of SN in our Galaxy occurring within the last several hundred years.  COMPTEL \citep{iyu94,sch00} has detected the 1157 keV $^{44}$Ti line from Cas A, the youngest known core-collapse SN remnant (age $\sim320$ yr) in our Galaxy.   \citet{vin01} reported the detection of the 68 and 78 keV lines from Cas A using the BeppoSax satellite, and later from INTEGRAL/IBIS as well \citep{vin05}.  The measured fluxes were consistent with that found previously by COMPTEL for the 1157 keV line.  Another detection of the 1157 keV line from the Vela region was reported by COMPTEL several years later \citep{iyu98}, however subsequent analysis \citep{sch00} has cast doubt on its validity.  The location of the new COMPTEL source was found to be coincident with a young SN remnant discovered independently and nearly simultaneously by the ROSAT satellite \citep{asc98}. 

\subsubsection{$^{60}Fe$}
\label{60fe}

$^{60}$Fe is also believed to be produced in core-collapse supernovae.  Its flux will accumulate from all of the Galactic supernovae occurring over its long half-life ($1.5 \times 10^{6}$ yr).  RHESSI \citep{smi04} and INTEGRAL/SPI \citep{har05} have both recently reported marginal (3-4 $\sigma$) detections of the 1173 and 1333 keV lines from $^{60}$Fe.  Theoretical models of Type II SN generally predict roughly the same amounts of $^{60}$Fe and $^{26}$Al \citep{lim03}.  However, the $^{60}$Fe/$^{26}$Al ratio reported by INTEGRAL/SPI is only $0.11 \pm 0.03$ \citep{har05}, and RHESSI finds a similar value.  The implication is that some other source of $^{26}$Al besides explosive nucleosynthesis in core-collapse supernovae is required.

\subsubsection{$^{56}$Co, $^{57}$Co} 
\label{57co}

SN1987A, a type II SN in the Large Magellanic Cloud, was the nearest supernova in the era of modern astronomy when instrumentation existed capable of detecting gamma-ray emission. Gamma-ray lines from  $^{56}$Co decay ($t_{1/2} = 77$ d, $E_{\gamma} =$ 847, 1238, 2599 keV) were first detected by the SMM satellite \citep{mat88} $\sim 6 $ months after the discovery of the SN.  The competition between the increasing transparency of the SN shell and the decay of $^{56}$Co results in a predicted gamma-ray light curve with a maximum $\sim 1$ yr after the initial explosion.   However gamma-ray emission was seen much earlier than expected, which argued for significant outward mixing of the $^{56}$Co which was initially produced deep inside the supernova shell.  Several high resolution germanium detectors were able to spectroscopically study the lines \citep{tee88,mah88,san88,tue90}.  Although not entirely self-consistent, the ensemble of observations showed broadened (FWHM = 10-15 keV) and redshifted lines that were consistent with emission from an expanding shell of velocity $\sim3000$ km/sec and substantial mixing.  

Supernova models have suggested that $^{57}$Co decay  ($t_{1/2} = 271$ d) might be the dominant energy source driving the later stages of the shell expansion of a Type II SN \citep{cla74,woo89}.  In 1991 the OSSE experiment on the Compton Gamma-Ray Observatory (CGRO) reported the detection of the 122 and 136 keV $^{57}$Co line complex from SN1987A \citep{kur92}.  OSSE did not have sufficient resolution to resolve the lines from each other.   The line flux implied that the ratio of $^{57}$Ni/$^{56}$Ni produced in the supernova was $\sim1.5$ times the solar system $^{57}$Fe/$^{56}$Fe abundance ratio.

\subsubsection{Solar Flares} 
\label{solar}

Energetic particles accelerated in solar flares can excite nuclei in the solar atmosphere that subsequently decay emitting gamma-rays.  These gamma-ray lines are useful diagnostics of the ambient conditions at the flare site and of the acceleration mechanism (for an extensive review see \citet{ram87}).  Extensive solar flare gamma-ray measurements have been carried out by the GRS experiment on the Solar Maximum Mission (see, for example \citet{for88}).  Gamma-ray lines from $\sim12$ of the most abundant elements below and including Fe have been identified.  Recently the RHESSI mission has provided high resolution spectroscopy using germanium detectors of solar-flare gamma-ray lines \citep{smi03}.  RHESSI has succeeded in resolving profiles of gamma-ray lines from Fe, Mg, Ne, Si, C and O.  The INTEGRAL instruments do not point at the sun, however for an intense solar flare there can be sufficient penetration of the shielding to allow useful gamma-ray line spectroscopy \citep{tat05,kie05}.

\section{Observations}
\label{obs}

\subsection{The INTEGRAL Mission}
\label{mission}
INTEGRAL is a powerful space-borne observatory of the European Space Agency (ESA) designed to make detailed measurements of the spatial and spectral distribution of celestial hard X-rays and gamma-rays.  For a detailed description of the mission see \citet{win03a}.  Launched in mid-Oct. 2002, INTEGRAL has operated successfully up to the present day.  The satellite was inserted into a highly eccentric orbit (10,000 km perigee, 160000 km apogee) with a period of 3 days.  In this orbit INTEGRAL spends very little time inside the radiation belts and avoids the damaging proton belts entirely.  INTEGRAL is operated as an observatory with $\sim$ 3/4 of it's time open to guest observers.  The remainder of the time (the so-called Core Program) is mainly dedicated to systematic deep exposures of the Galactic Center region and periodic scans of the Galactic Plane.   

\subsection{The INTEGRAL Spectrometer (SPI)}
\label{spi}
The Spectrometer for INTEGRAL (SPI) is one of the two major instruments on the INTEGRAL mission \citep{ved03}. SPI employs an array of 19 high-resolution cooled germanium detectors in a close-packed hexagonal array. It covers the energy range 20 - 8000 keV, over which the energy resolution increases from $\sim 2$ to 8 keV FWHM. The detectors are cryogenically cooled by two dual Stirling cycle refrigerators to a nominal operating temperature of 86K. In addition to providing a precise determination of the photon energy, the Ge detector array also acts as an imaging device in conjunction with a tungsten coded-aperture mask located 1.7 m from the detector plane. Images are formed by deconvolving the shadow pattern cast by the mask on the detector plane. The fully-coded field-of-view of the telescope is $16 \degr$ and the angular resolution is $\sim 3 \degr$ FWHM. The Ge detector array is surrounded by an active anti-coincidence shield/collimator made of bismuth germanate (BGO) scintillator. This dense hi-Z material is ideally suited for suppressing the strong flux of background radiation generated in both the SPI instrument and in the spacecraft.  The active collimator defines a field-of-view of $\sim25 \degr$ FWHM.

\subsection{Data Selection}
\label{data_sel}
\begin{figure}
\plotone{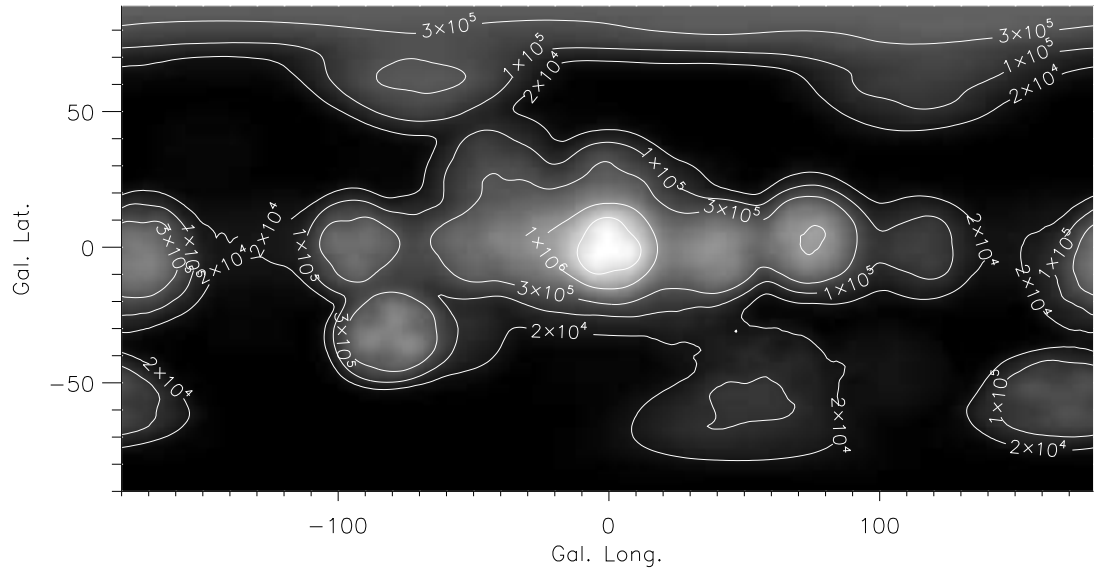}
\caption{INTEGRAL/SPI Exposure map of the Galaxy for the first year of observations. Contours are labeled with the exposure in seconds.  The deepest exposures are in the Galactic Center, Cygnus, Vela and SN198A regions.  This map assumes the so-called "light-bucket" method of analysis (see Section~\ref{diff_src}   for a detailed discussion). The characteristic ÒpinwheelÓ patterns in some regions are due to modulation of the response by the SPI coded-aperture mask.  The map was generated using the SPI response function at 500 keV.}
\label{exp_map}
\end{figure}

The data used in this investigation span the period from 2002 December to 2003 December.  All INTEGRAL data becomes public after $\sim$ one year so we have been able to use essentially all of the data from this period.   The data have been carefully screened to remove bad periods due to enhanced solar activity, trapped radiation and noise.  $\sim12$ hr of data near perigee each 3 day orbit is not usable.  In addition there were three SPI anneallings during this period resulting in the loss of $\sim$ 35 days of data.  After all data selection criteria were applied the total remaining usable time in the first year was 12.3 Msec.  A large fraction of the total observing time was spent in or near the Galactic Plane.    3.6 Msec were spent in the Galactic Center region (within $30\degr$ of the center)  and 9.6 Msec were within $\pm 20 \degr$ of the Galactic Plane.  An exposure map of the data used in this paper is shown in Fig.~\ref{exp_map}.  The map was generated using the SPI response function at 500 keV.

During nearly all observations the direction of pointing is stepped through a "dithering" pattern.  This provides additional information that is useful (essential in the case of SPI) for both determining the background and producing high quality images.  For most point source observations a square $5  \times 5$ pattern with $2 \degr$ step size was used.  The duration of each pointing was typically $\sim 2000$ sec.  For the Core Program deep exposure of the Galactic Center a raster scan was performed which typically covered $\pm 30 \degr$ in Galactic longitude and $\pm 20 \degr$ in Galactic latitude with step size and duration similar to that for point sources.   Our total data set comprises 6194 of these pointings.

\subsection{Instrument Background}
\label{inst_bck}
\begin{figure}
\plotone{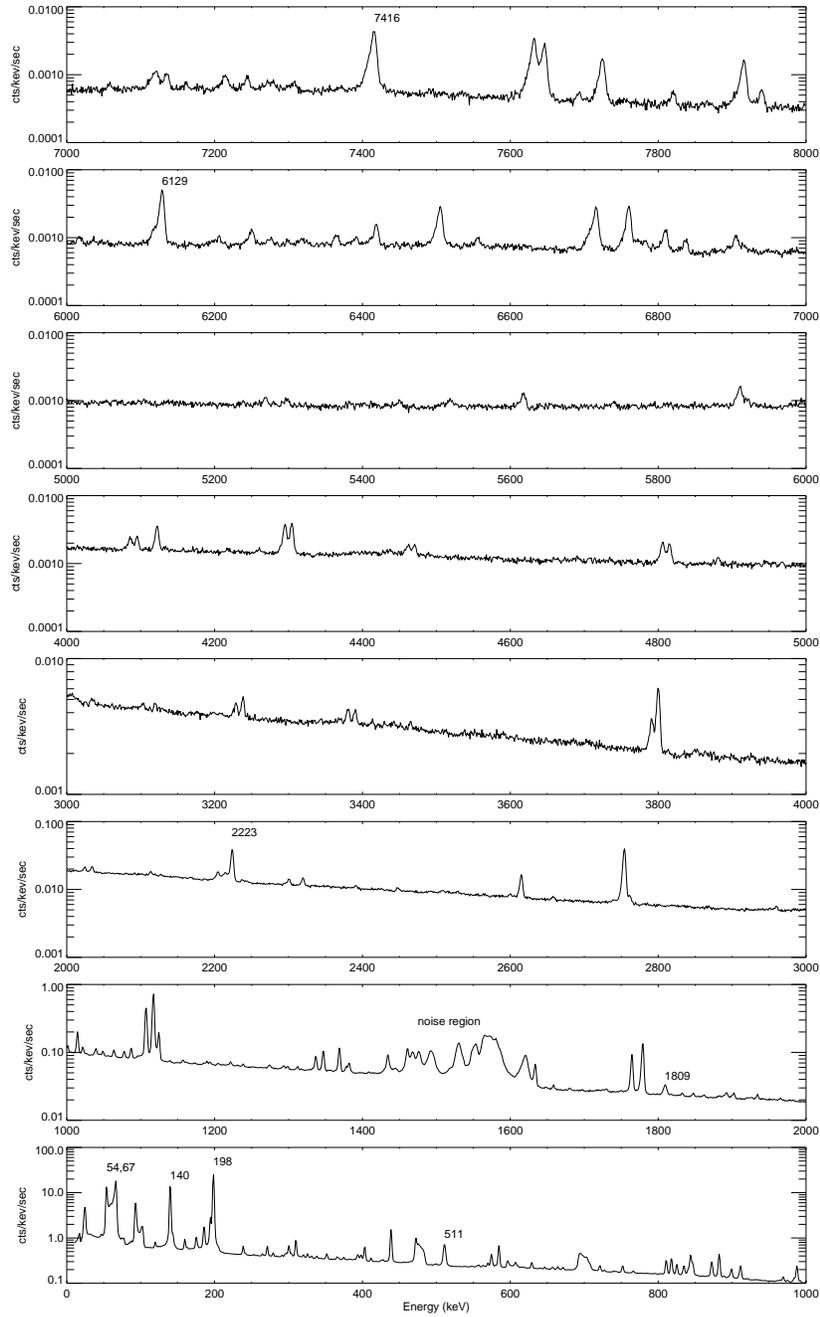}
\caption{INTEGRAL/SPI instrumental background spectrum.  Strong background lines and those at astrophysically interesting energies are labeled with their energy in keV.  Also labeled is the complex instrumental noise feature in the 1400-1600 keV region. }
\label{bckgnd}
\end{figure}

Background is primarily produced in SPI through the interaction of cosmic rays with the telescope and with surrounding material in the spacecraft (for detailed discussions see \citet{jea03,wei03,tee04}). The diffuse cosmic background also contributes mainly at energies below $\sim 200$ keV.  Cosmic rays colliding with nuclei in the detectors and surrounding material can produce many different isotopes.  Many of these are created in excited unstable states which subsequently decay with the emission of gamma-rays (and other particles as well).  Gamma-rays are produced at (nearly) discrete energies corresponding to the transitions between various nuclear levels. This leads to the complex multi-line SPI background spectrum shown in Fig.~\ref{bckgnd}.  For a nearly complete identification of the SPI background lines see \citet{wei03}.  The strongest background  lines as well as certain lines that fall at astrophysically interesting energies (e.g. 511 keV, 1809 keV) are labeled with their energies in keV.  Below $\sim 200$ keV the spectrum is dominated by very strong lines (54, 67, 139, 198 keV) that are primarily emitted by long-lived isomeric states which are not rejected by the SPI active anti-coincidence shield. These lines account for $\sim 30 \%$ of the total SPI background counting rate.  There is a strong background line at 511 keV which is particularly problematic for the astrophysical 511 keV analysis.  A weaker background feature due to a complex of lines is present at 1809 keV which must also be taken into account.  Above 6 MeV a number of  lines are seen corresponding to the highest energy levels of the isotopes of Ge.   Here also, one sees the effects of instrumental line broadening which increases monotonically with energy.  The complex feature between $\sim 1400$ and 1600 keV is an instrument artifact most probably arising from very large ($> 1$ GeV) energy losses in the Ge detectors due to nuclear spallation and/or heavy cosmic rays.  We have excluded this region from our analysis.

\begin{figure}
\plotone{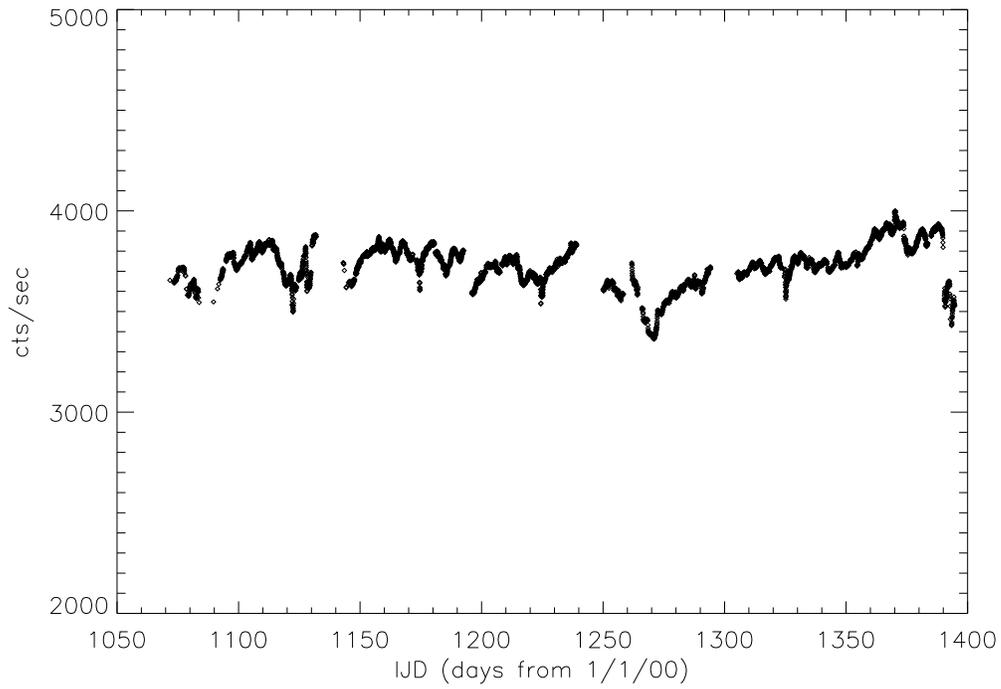}
\caption{INTEGRAL/SPI germanium detector saturating event rate.  This shows the typical temporal behavior of the SPI background.  Periods of enhanced flux due to solar flares and electrons that have escaped the earth's magnetosphere have been removed.}
\label{gedsat}
\end{figure}

The cosmic rays producing the SPI background vary in intensity due to modulation by the solar magnetic field.  The SPI background responds in a complex way to these variations.  In prior analyses \citep{jea03,tee04} the rate of events saturating the Ge detectors (GEDSAT) was found to be a good tracer of the SPI 511 keV background.  It is dominated by energetic particles (i.e. cosmic rays) with E $> 8$ MeV.  This rate is plotted in Fig.~\ref{gedsat}.  During the one-year period under study the total variation is $ \sim 10 \%$.  Over longer time intervals larger variations are expected due to the 11 yr period of the solar cycle.   Most of the variability is on time scales longer than a few days.  We will return to this point later as it is crucially important to our methods of background subtraction.  

In addition to the GEDSAT rate a variety of possible choices of background tracers are available.  For example, there are other detector rates, shield rates, and background line rates.    We found in our analysis that residuals at the locations of strong background lines (see Section~\ref{large_scl}) were particularly problematic.  Hence we chose to use the strongest background line (198 keV) as our primary background tracer.  Ideally one might like to tailer the tracer to the particular line under investigation.  Given the wide range of lines searched for in this study, that was not practical.  

SPI is a sensitive instrument and with long accumulation times is capable, in principle, of detecting very weak gamma-ray line fluxes.  However, at its limiting sensitivity (for, say, T~=~$10^{6}$ sec), the typical signal-to-background ratio is only a fraction of a percent.  Using a tracer we can normalize  and subtract the background.  However, the subtraction is not perfect.  In the simplest case the intensity of a particular background line is determined by a convolution of the cosmic-ray spectrum with the energy-dependent cross section for the production of that line.  The shape of the cosmic-ray spectrum changes with changing solar modulation, which means that there is not a simple linear dependence between the cosmic-ray and instrument background rates.  In addition a significant number of background-producing  isotopes have long ($> 1$ day) half-lives .  This again means that the SPI background will not simply track the cosmic-ray intensity.   Our approach will be to choose the background dataset near in time to the source observation (all data taken within 20 days of the source observation and farther way than $20\degr$).  We will see later that this can very effectively suppress the background-subtraction residuals.  It will be necessary to trade off suppression of residuals against loss of sensitivity due to the reduced availability of background data.    We also found that the standard procedure for gain correction, i.e. performed only once per orbit, was not adequate.  There were times when gain variations within an orbit could produce significant residuals in the background-subtracted spectra.  We therefore performed a secondary gain correction on  a pointing-by-pointing basis which improved the background subtraction significantly.  We used a linear interpolation between residuals at four background lines (138,  439, 2754 and 6129 keV).  We found that a single simple linear correction was adequate even though it crossed the boundary between the two different SPI gain regions.  The secondary gain correction was tiny and only significant for a few of the orbits, where probably there were some temperature drifts.  The main purpose of the correction was to suppress residuals at strong background lines, which it successfully did.  The background-subtracted spectra obtained during those orbits were all visually examined and no significant residuals were found, including the 1765 keV line.

\section{Data Analysis}
\label{data_anal}
 
\subsection{General Search Procedure}
\label{sea_proc}
The line search procedures used in this study varied according to whether or not the assumed sources were diffuse or point-like, steady or variable.   For diffuse emission they also varied with the scale of the assumed source.  However,  the following elements were common to all procedures.  First, a background-subtracted spectrum was created by using close-in-time background spectra and scaling them in intensity according to the 198 keV line intensity, as explained above.  This is the most crucial step.  As was seen in Fig.~\ref{bckgnd} there are numerous strong background lines.  These lines vary in strength and relative intensity.  Also, the line profiles change with time due to radiation damage.   All of these effects can produce residuals in the background-subtracted spectra that can masquerade as astrophysical lines.  A number of different background subtraction methods were investigated.  The most effective method and final choice was the application of a time-dependent background selection criterion as discussed in the preceding section.

\begin{figure}
\plotone{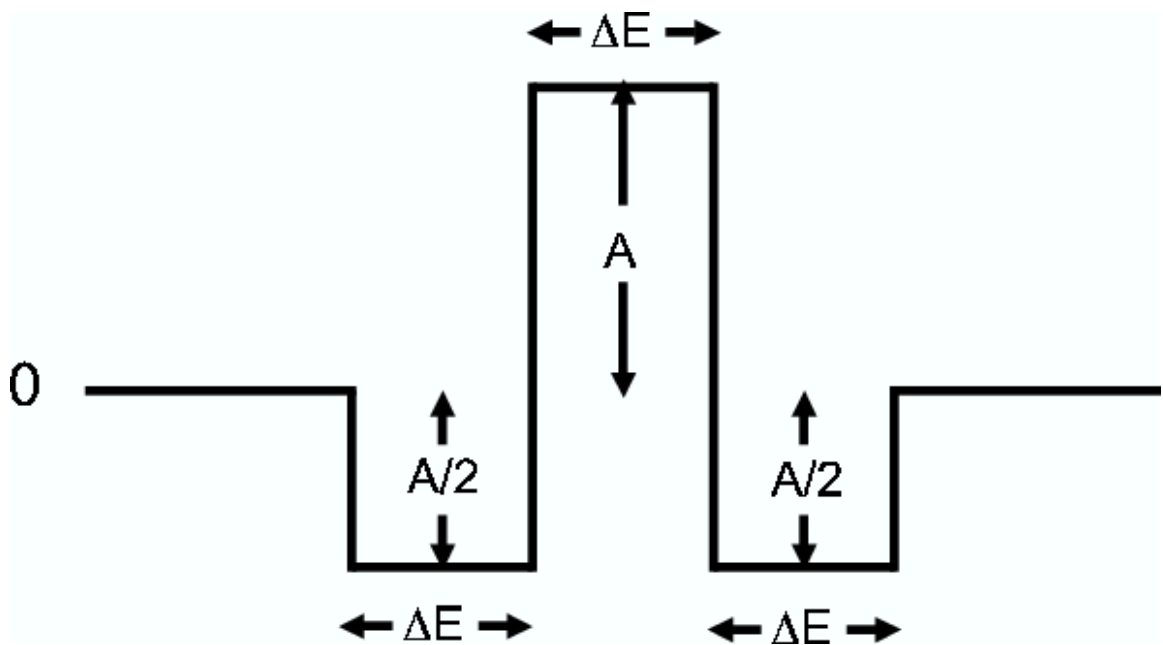}
\caption{Convolution template used in line search.}
\label{template}
\end{figure}

\begin{figure}
\plotone{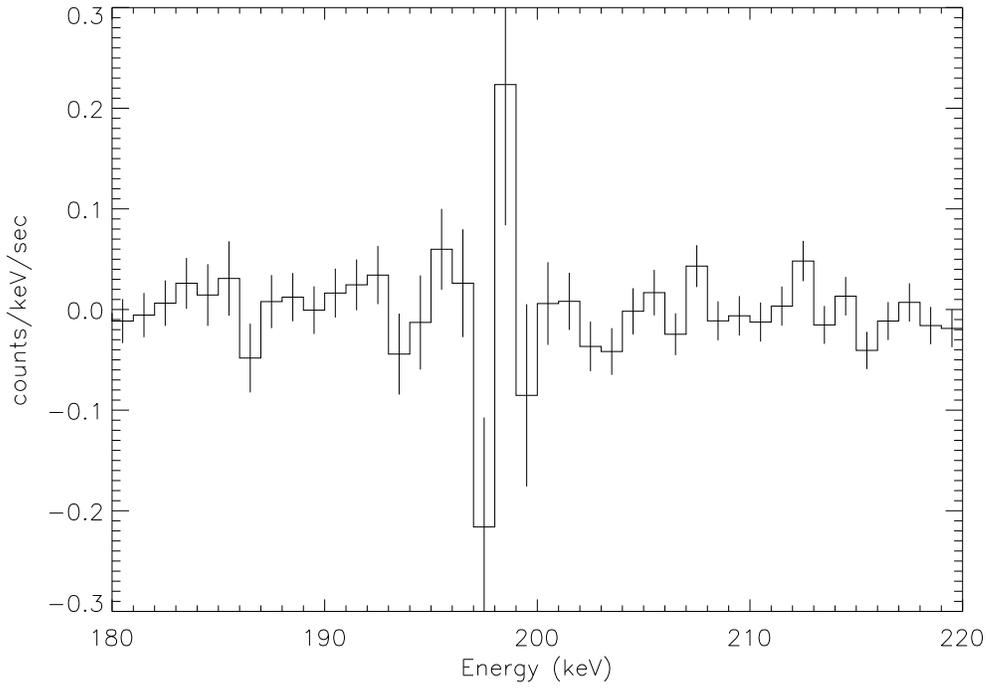}
\caption{Example of bipolar residual from strong background line at 198 keV due either to line broadening or gain shift.}
\label{bipolar_res}
\end{figure}

\begin{figure}
\plotone{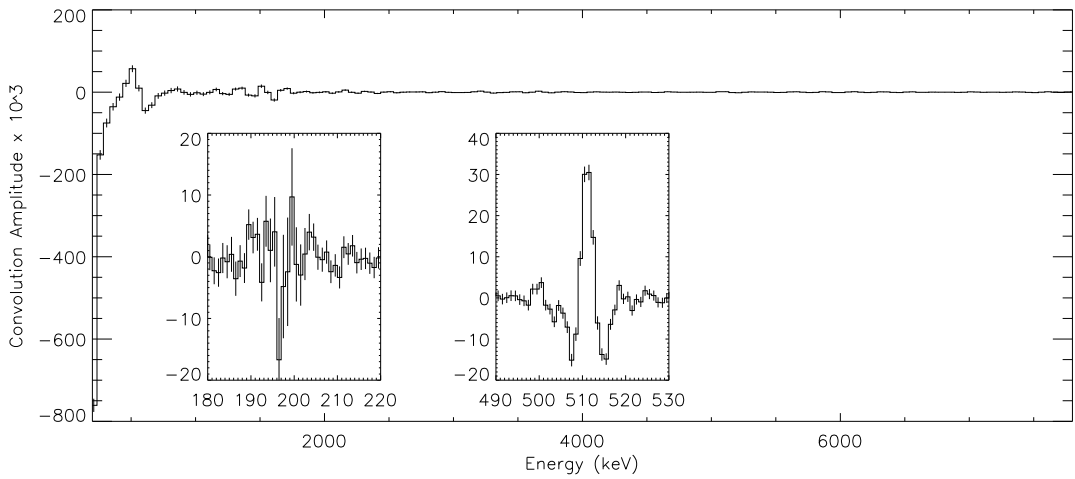}
\caption{Convolved spectra from Galactic Center data set.  Large panel shows entire range using 100 keV convolultion template.  Some excess is seen in the 1.4-1.6 MeV noise region which has been excluded from the analysis.   Left inset panel: region around strong 198 keV background line (3 keV template).  Right inset panel: region around 511 keV line (3 keV template). }
\label{convol}
\end{figure}

The background-subtracted spectra were all then run through the same search procedure.  The spectra were convolved with the template shown in Fig.~\ref{template}.  The effect of this convolution is to subtract the local continuum from the line, which is necessary when there is a significant continuum from the source.  It also removes any residual continuum due to imperfect background subtraction.  The convolution was repeated while varying the width $\Delta$E of the convolution kernel over the range 3-1000 keV in $20\%$ increments.  A detection is defined as a single positive deviation from zero that exceeds the threshold for that analysis (see below for threshold definition). Various quality checks were included the software.  For example, bipolar residuals, which can be produced when the width of a background line varies or the gain changes, were rejected.  An example of such a feature is given in Fig.~\ref{bipolar_res}.  In Fig.~\ref{convol} we show an example of a convolved spectrum.  The spectrum is of the Galactic Center region of the type used in the small-to-medium scale analysis (Section~\ref{smed_scl}).  The large panel contains the spectrum convolved with a 100 keV template.  Regions within two kernel widths (200 keV in this case) of the spectrum upper and lower boundaries are excluded since detecting a line feature requires knowledge of the continuum on either side of the line.   A negative residual is evident below $\sim 300$ keV which degrades the sensitivity for line detection in this region.  This unusual case is due to the strong continuum from the Galactic Center region.  Also shown are two insets giving magnified views with finer binnings of two regions of the plot.  The left inset is the region around the strong 198 keV background line with 3 keV binning.  It displays a complex structure due to tiny changes in the line shape and position, but none of the positive points exceeds the $5 \sigma$ detection threshold.  The right panel shows the region of the 511 keV line, also with 3 keV binning, which is clearly detected.  

The volume of phase space and effectiveness of residual suppression varied with the different search methods.  Hence it was necessary to vary the detection thresholds to assure an acceptably small number of false positives.  These thresholds were: large-scale search, $4 \sigma$; small-to-medium scale search, $5 \sigma$; rapidly variable search, $4 \sigma$; point source search, $3.5 \sigma$.  As an aid, software tools were created that allowed the rapid viewing of large numbers of spectra to verify the procedures and examine candidate lines.

\subsection{Diffuse Sources}
\label{diff_src}
For the diffuse source search we use the so-called "light bucket" approach.  In this method all 19 SPI detectors were summed, effectively creating a simple collimated instrument whose field-of-view ($\sim 25 \degr$ FWHM) is determined by the SPI active BGO collimator.  A summed raw spectrum was created for each of the 6194 pointings.  From these spectra we created two different kinds of background-subtracted spectra described in the following sections.

\subsubsection{Large Scale ($\gtrsim20 \degr$)}  
\label{large_scl}

By large-scale we mean spectra accumulated over large scale regions, specifically the Galactic Center (within $13 \degr$ or $30 \degr$ of the center) and Galactic Plane (within $\pm 20 \degr$ of the plane excluding the central region).   The available source pointings were accumulated into daily spectra  and background pointings for each source spectrum were chosen according to the following criteria: 1) the SPI axis direction is not in the source region and 2) the time interval between source and background pointings was $ < 20$ days.  The choice of 20 days was based on the trade-off discussed earlier between residuals in the spectrum and loss of sensitivity due to the reduction in the available background data.  Basically, we used the largest possible time interval that produced an acceptably small number of false positives.  Background spectra were normalized to source spectra before subtraction using the 198 keV instrumental line, which provided the best cancellation of the strong background lines.  Finally an average was taken over all of the Galactic Center daily spectra and separately over the Galactic Plane daily spectra.  The daily spectra may have common background pointings and are thus not necessarily independent.  This was taken into account  in the error and sensitivity calculations. 

\begin{figure}
\plotone{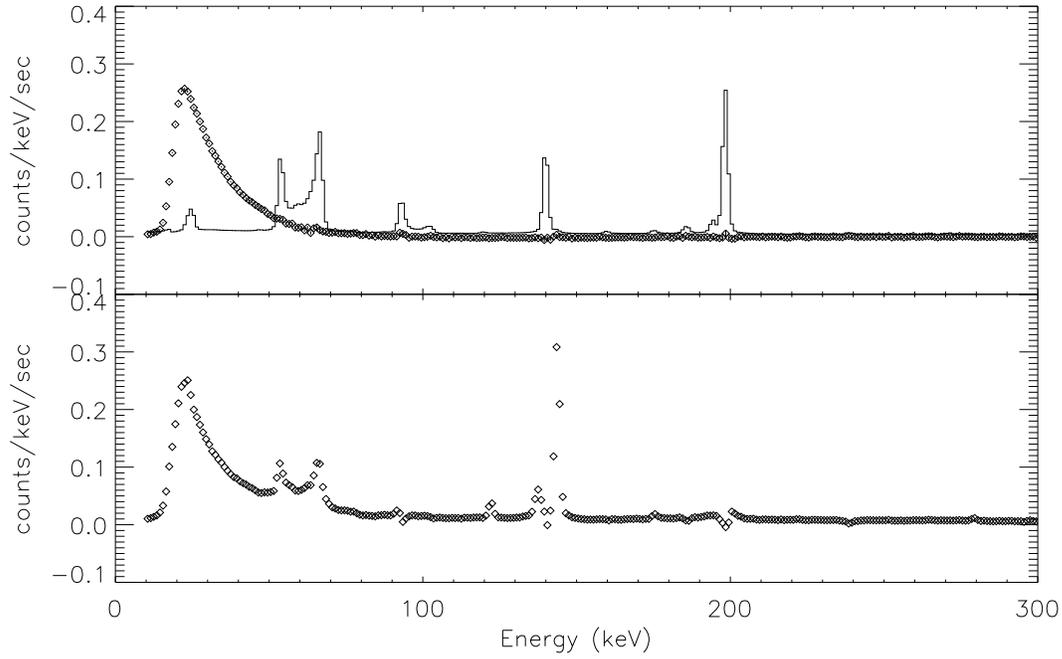}
\caption{Galactic Center - Off-Center difference spectra. Upper panel: 20 day background selection criterion.   Lower panel: no background selection criterion.   Dramatic suppression of background line residuals is evident with 20 day selection criterion.  Solid line in upper panel is $1 \%$ of total background spectrum.  Residuals at the locations of the strong lines are $\lesssim 0.1 \%$ of the raw background.}
\label{spdiff_plot}
\end{figure}

Fig.~\ref{spdiff_plot} shows large-scale difference spectra formed by subtracting all the off-center data ($> 30 \degr$ from the center) from the on-center data.  Some overlap may exist in the fields covered by the source ($\theta < \theta_{lim}$) and background ($\theta > \theta_{lim}$) spectra which could cause some cancellation of signal especially in the case of the $\theta_{lim} = 13 \degr$ data set (instrument collimation FWHM ~24deg).  This was deemed an appropriate sacrifice in order to maximize the amount of close-in-time data available for background subtraction.  The 20-300 keV region containing the strongest background lines is plotted.  The individual data points are the background-subtracted count rates.  In the upper panel the background (off-center data) has been selected as described in the preceding paragraph using a value of 20 days for the maximum allowed time difference between source and background.  In the lower panel no background selection has been applied and strong residuals are evident.   The 198 keV line has been used for normalization so that its net residual is zero.  However, there is a significant bipolar feature due to the changing shape of the line (a radiation damage effect) and possible uncorrected gain shifts.  Other bipolar residuals are seen at  93 keV and 139 keV.  The strongest residual is from the 143 keV background line (a rather weak one).  The reason for it's prominence is that it is produced by the decay of $^{57}$Co with a half-life of 271 days.  The strength of this line will build up over this period and it will not track the variations of the other lines.  This spectrum illustrates the critical importance of proper selection of background data.  Also shown in the upper panel is the raw unsubtracted spectrum multiplied by 0.01.  The effectiveness of the line cancellation is evident - the residuals at the locations of the strongest lines are $\lesssim 0.1 \%$.

\subsubsection{Small-to-Medium Scale ($\lesssim20 \degr$)}
\label{smed_scl}

To search for smaller scale diffuse emission a series of uniformly-spaced grid points on $18 \degr$ centers in Galactic coordinates was defined.  For each point on the grid source data were taken to be all pointings with $12 \degr$ (the half-response) of that point.  As before, the available source pointings were binned in 1 day intervals thus creating a set of daily source spectra for each grid point.  Background spectra for each grid point source spectrum were chosen as follows: 1) the angular distance from the grid point was $> 20 \degr$ and 2) the time interval between source and background pointings was $ < 3$ days.   In this analysis it was possible to use a smaller 3 day limit since background pointings were generally more readily available than they were for the large-scale analysis.  Again, the 198 keV line was used to normalize the amplitude of the background spectrum to the source spectrum.  Also, as before, an accumulated average spectrum was calculated for each grid point. Both the individual daily spectra and the average spectra were searched for lines following the procedure of Section~\ref{sea_proc}.  Although intended primarily for the search for diffuse lines, this procedure could also detect lines from point sources in the neighborhood of each grid point if they were present.  It could conceivably catch point sources missed in the SPIROS-based analysis described in Section~\ref{pt_src}.

\subsection{Point Sources}
\label{pt_src}

The SPI Bright Source Catalog\footnote{\url{http://heasarc.gsfc.nasa.gov/docs/integral/bright\_sources.html}} containing 159 sources was used as a primary input for the search for gamma-ray lines from point sources.    The SPI standard analysis pipeline software was used [OSA Version 5.0 \citep{die03a,wal04}].  Point source spectra were created by the SPIROS program  [SPI Iterative Removal of Sources \citep{dub05}].  The program initially creates a correlation map.  The brightest source is identified, convolved with the response function and subtracted from the data.  This process is repeated until no further significant sources remain.  In this manner the effects of ghost images from bright sources in the field-of-view are suppressed or eliminated.  The spectral binning (1 keV,  20-2000 keV; 10 keV, 2000-8000 keV) has been  chosen to ensure that there are sufficient counts per bin to preserve near-gaussian statistics (and therefore a linear fitting procedure) while maintaining the finest possible resolution to allow searching for the narrowest possible lines.   In addition to the per orbit spectra a cumulative average spectrum over all the available data was determined.  Finally, we also have created stacked spectra over generic types of sources (LMXB, HMXB, pulsar, SNR, Seyfert).  In the per-orbit spectra the background is fit at the same time as the source, which means that it comes from the same orbit as the source.  The background subtraction is therefore immune to most of the long-term effects described in Section~\ref{inst_bck}.  This is true for the point source spectra averaged over time and over source type as well.

\begin{figure}

\plotone{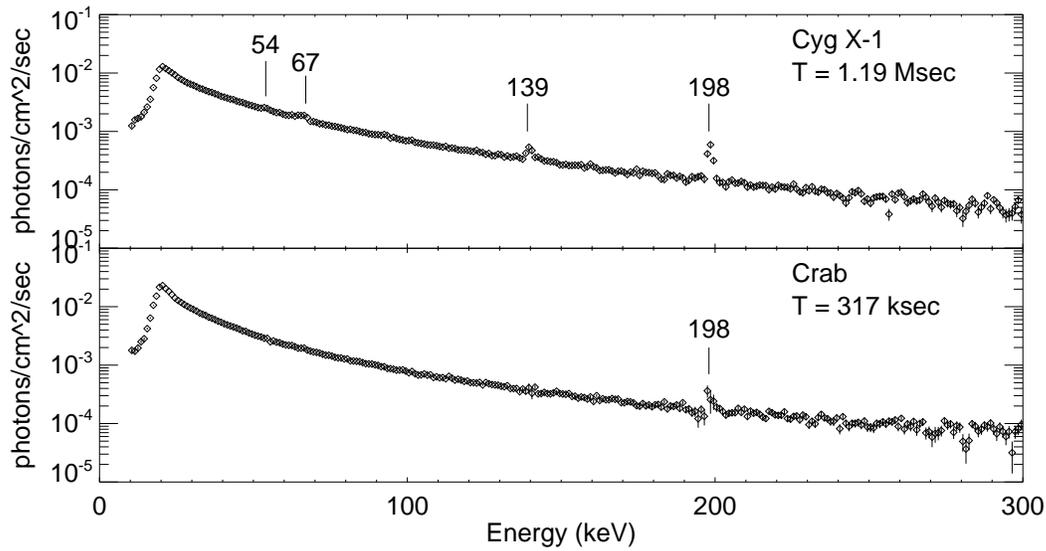}
\caption{SPIROS spectra of the Crab and Cyg X-1 in the 20-300 keV range with 1 keV binning. T = spectrum integration time.  Labelled lines are residuals of strong instrumental background lines. Error bars are statistical only. }
\label{crab_cyg}
\end{figure}

In Fig.~\ref{crab_cyg} we show spectra from the Crab and Cyg X-1.   These  are the principal INTEGRAL calibration targets and  were observed for long periods of time.  We have chosen these long accumulations since they best illustrate the effects of residuals due to the imperfect removal of instrumental background lines.  Again the spectra are plotted with 1 keV binning in the 20-300 keV interval, which is the region where the strongest background lines are located.  For the Crab spectrum with an integration time of 317 ksec there is a line-like residual at 198 keV, the energy of the strongest background line.  The error bars are large so that the significance of this feature is not great.   No other significant residuals appear in the Crab spectrum.  For Cyg X-1 with a longer integration time (1.19 Msec) more residuals appear at 54, 67, 139 and 198 keV.   These spectra illustrate the limiting factor of the line search.  Spectra of durations up to a few hundred ksec are relatively systematics free.  For spectra of longer duration the number of line-like artifacts increases.  We can (and have) searched these spectra, but we needed to perform an additional screening to remove the residuals at the locations of the strongest background lines.  This was done by hand by simply checking each detection to see if it coincided with a strong background line.  Generally speaking the number of false positives was small and restricted to the few strongest background lines.  This does imply a degradation in sensitivity for long spectral accumulations in the vicinity of the strong background lines.

\section{Results}

\subsection{Diffuse Sources}
\label{res_diff}

\subsubsection{Large Scale Diffuse Emission}
\label{ls_res}

We created three different large-scale background-subtracted spectra with parameters summarized in  Table~\ref{ls_parms}.  Our standard line search algorithm was applied (see Section ~\ref{sea_proc}) to these three spectra with a $4 \sigma$ threshold.  Several line-like residuals were found at the positions of strong background lines, but no new lines appeared at other energies.  The 511 keV and 1809 keV lines were detected at the expected fluxes and significances.  For large-scale diffuse emission the flux sensitivities depend on the assumed shape of the spatial distribution.  We calculated sensitivities for five different distributions summarized in Table~\ref{spatial_distr}.  The $10 \degr$ gaussian (I) was chosen since it characterizes the known distribution of the bulge component of the $e^{+}$ - $e^{-}$ annihilation radiation.  The flat circular distribution (II) was included as an unbiased sample of the central radian of the galaxy.  The flat longitudinal distribution (III) is used as unbiased sample of the Galactic Plane.  The Dirbe $60 \micron$ distribution (IV)was included as representative of the massive star distribution in our Galaxy since nucleosynthesis in massive stars is a likely source of radioactive nuclei.  It is one of the best matches to the observed $^{26}$Al 1809 keV spatial distribution  \citep{kno99}.  Finally, the Egret 30-100 MeV distribution (V) was used as representative of the cosmic-ray + interstellar gas distribution.  In Fig.~\ref{ls_sens} we show the sensitivity as a function of energy for the Galactic center data set ($13 \degr$ radius)  and $10 \degr$ gaussian case (GC13,I).  The curves are parametrized by line width (FWHM).  The large-scale sensitivity results are summarized for all the appropriate combinations of data sets and flux distributions in Table~\ref{sens_summ_lar}.  Narrow-line sensitivities are given at several selected energies.   Also given for each grid point is a scale factor by which to multiply the curves of Fig.~\ref{ls_sens} in order to obtain sensitivities at other energies and line widths.  

\subsubsection{Small-to-Medium Scale Diffuse Emission}
\label{smed_res}

\begin{figure}
\plotone{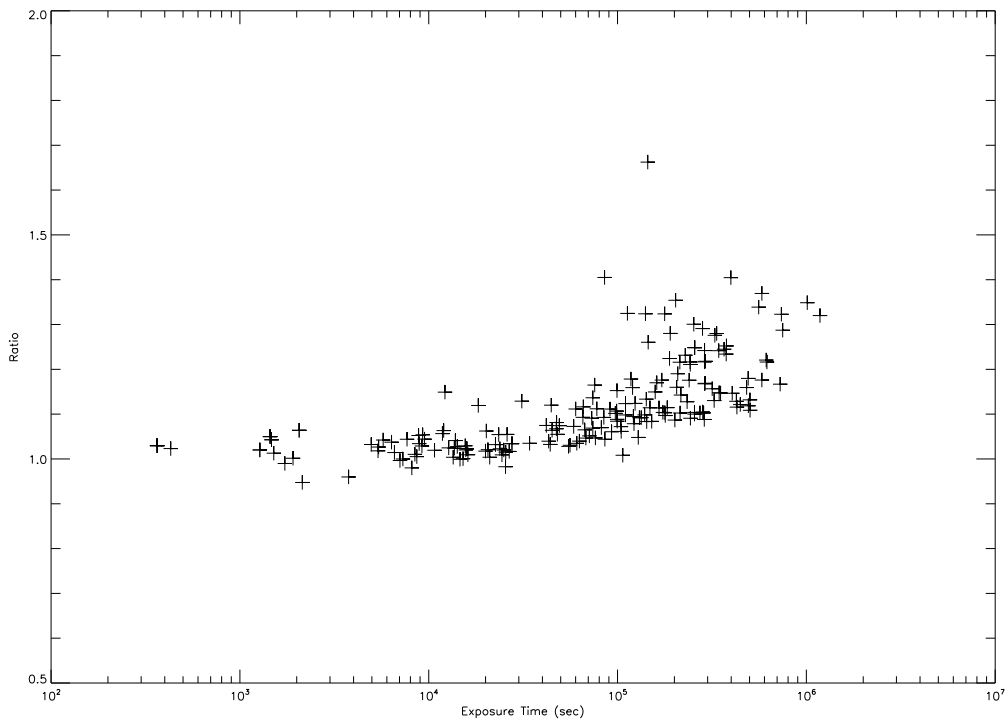}
\caption{Energy-averaged ratio of standard deviation to median error of data points of difference spectra (similar to Fig.~\ref{spdiff_plot}) vs. spectrum accumulation time.  Spectra with long accumulation times display larger ratios reflecting increasing importance of systematic residuals.}
\label{stat_ratio}
\end{figure}

Daily background-subtracted spectra were created for each Galactic grid point as discussed in Section~\ref{smed_scl}.  We tested the spectra for well-behaved errors by calculating the standard deviations the residuals over several broad energy bands in the 1000-8000 keV range.  This high energy range was chosen to avoid non-zero residuals due to strong sources.  For each energy range the ratio of standard deviation to mean statistical error was calculated.  Fig.~\ref{stat_ratio} shows the energy-averaged ratios  for each Galactic grid point plotted as a function of the total accumulation time.   For purely statistical errors this ratio should be close to unity independent of accumulation time.  For accumulation times $\gtrsim10^{5}$ sec a moderate departure from unity is seen.  This is interpreted as due to the decreasing statistical error becoming comparable to the systematic error due to imperfect background subtraction.  This effect was part of our motivation for choosing  a threshold of $5 \sigma$ for the small-to medium scale analysis.

\begin{figure}
\plotone{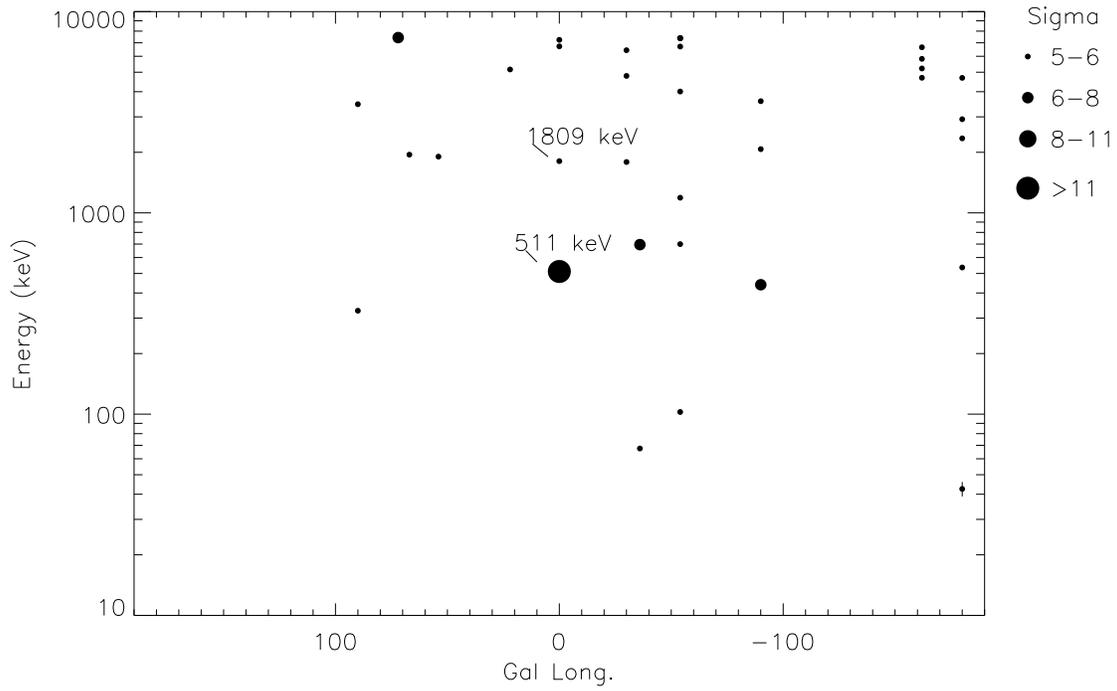}
\caption{Results of small-to-medium scale diffuse line search using program described in  Section ~\ref{sea_proc}.  These results are derived from the accumulated average of the daily spectra for each Galactic grid point. Horizontal axis is Galactic longitude.  Galactic latitude is suppressed in order to allow creation of 2D plot.  Vertical axis is line energy.  Vertical bars through data points show line width (not error).  The $^{26}$Al (1809 keV) and $e^{+}$ - $e^{-}$  annihilation (511 keV) lines are labelled. These lines are detected by our program at the correct spatial location, energy and intensity.  }
\label{sms_lb_plot}
\end{figure}

Fig.~\ref{sms_lb_plot} is an example of the sort of plot we used to scan the data for interesting line candidates.  The vertical axis is the line energy and in this case the horizontal axis is the Galactic longitude of the grid point.  (Here the Galactic latitude is suppressed to allow creation of a 2D plot.)  The vertical bars seen in some points are the line widths (not errors).  The symbol size reflects the significance of the line detection.  The  511 keV  and 1809 keV lines are identified.  They are detected at the correct spatial location, energy, intensity and significance, which validates our search procedure.  The 511 keV line is the most significant of all those appearing in the plot.  1809 keV emission has been previously detected over a broad region of the Galactic Plane \citep{kno99,die06}. Our search detects it only at two grid points near the GC, which is to be expected since the flux is weak away from the GC, and more sophisticated analysis techniques are required.  Three lines (439 keV, 696 keV, 7430 keV) are detected at $> 6$ sigma.  However, all of them coincide with strong instrumental lines and are from grid points with long accumulation times ($> 3 \times 10^5$ sec).  A number of other line candidates are seen, however they are all near the $5 \sigma$ threshold.  We have individually examined all of these lines, and we interpret them as spill-over from our $5 \sigma$ threshold due to imperfect background subtraction.    

We find no lines at astrophysically interesting energies from individual grid points such as would be expected from small ($\lesssim 20 \degr$) emission regions or point sources. Nor do we find any clustering of candidates in energy-longitude space, such as might be expected from a region of diffuse emission more extended in longitude.

In Table~\ref{sens_summ_smed} we summarize the SPI sensitivities at each of our Galactic grid points.  We show only those points for which there was a significant exposure (T   $> 10^{4}$ sec).  As in Table~\ref{sens_summ_lar} we give the narrow-line sensitivities at several selected energies including 511 keV and the scale factor to apply to Fig.~\ref{ls_sens}  in determining the sensitivities at other energies and line widths.

\subsection{Point Sources}
\label{res_pt}
Fig.~\ref{crab_sens} shows the line sensitivity for the Crab as a function of energy for four different line widths.  This plot is to be used as a template (similar to Fig.~\ref{ls_sens})  for determining the sensitivities of other point sources.  A total of 2661 spectra from 143 different sources were created and searched. Tables~\ref{sens_lmxb} to \ref{sens_misc} summarize the results.  The tables are organized by type of source (LMXB, HMXB, Pulsar, Quasar, SNR).  Where it was relevant an average spectrum over all sources within the given type was created and searched for lines.   The Seyferts (Table~\ref{sens_seyfert}) span a range in redshifts of 0.001 - 0.063.   When the average Seyfert spectrum was created a redshift correction was applied.  No average spectrum for quasars was determined as their redshifts cover a wide range.  As before, narrow ($< 3$ keV) line sensitivities are given for four selected energies and a scale factor (to multiply Fig.~\ref{crab_sens}) is included for determination of sensitivities at other line energies and widths.  The tables show  only the results for the total accumulated spectra for each source.  We also created and searched individual spectra for each revolution (3 day period) in which the source was observed.   Finally we selected observations for each source in which the source was detected by SPI ($> 6\sigma$) in the 20-40 keV range and created and searched spectra from this subset.  As before, both source-averaged and individual spectra were examined.  The results are too numerous to include here.  Suffice it to say that no lines were detected in any of these spectra. 
 
\subsection{Rapidly Variable Sources}
\label{rapid_var}
\begin{figure}
\plotone{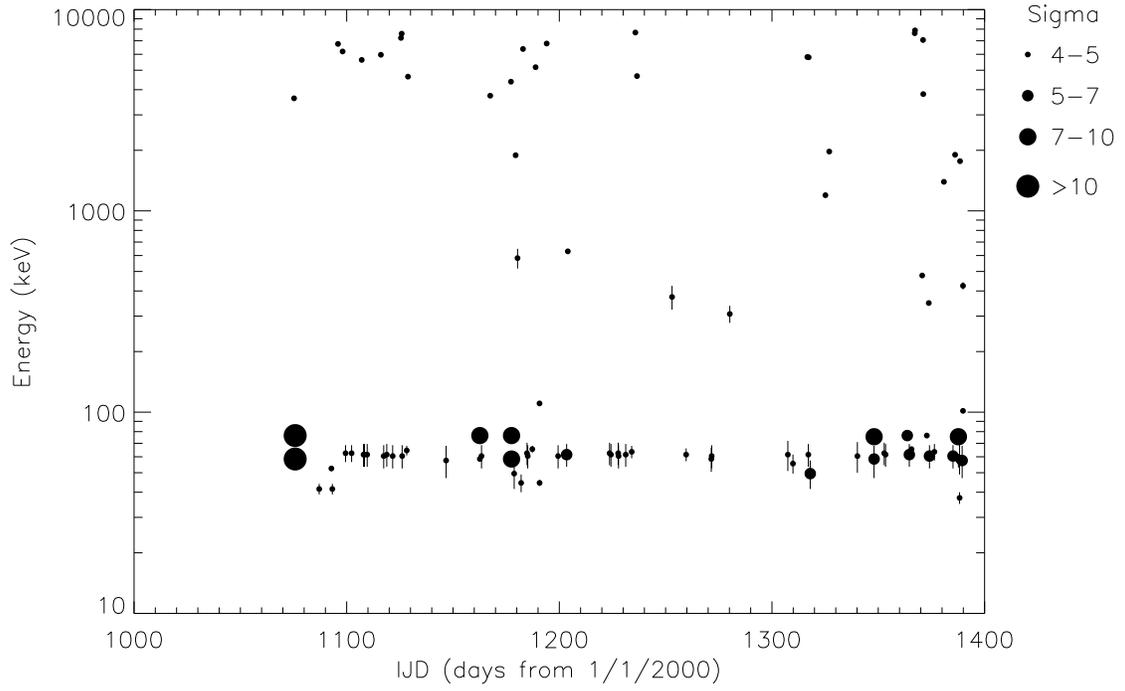}
\caption{Results of rapidly variable line search using program described in  Section ~\ref{sea_proc}.  Vertical axis is line energy.  Vertical bars through data points show line width (not error).  The strong line pairs at 58 and 76 keV are from occasional bursts of magnetospheric electrons.}
\label{ls_scw_plot}
\end{figure}

The line search discussed in Section~\ref{smed_res} was sensitive to sources which vary on time scales $\gtrsim 1$ day.  We also conducted a search for sources that vary on shorter time scales.  A convenient time interval is that corresponding to a single pointing (typically 1000-2000 sec duration).  The data and standard software are organized in a manner that makes such a search relatively easy.  Spectra were created for each pointing and screened as before for solar and magnetospheric particle contamination.  A total of 6004 spectra were produced.  Background was simply taken to be the 1-day average spectrum from the day in which the pointing occurred.   To maintain well behaved errors, we found it necessary to rebin the $> 2000$ keV part of the spectrum into 10 keV bins (as was done for the point source spectra).  We are therefore less sensitive to lines of width $< 10$ keV having energy $> 2000$ keV, but the $> 2000$ keV instrumental resolution is not too much smaller than 10 keV, so not much is lost.  The same line search algorithm described in Section ~\ref{sea_proc} was used to search all of the spectra.  Fig.~\ref{ls_scw_plot} shows the results of the search.  It is similar to Fig.~\ref{sms_lb_plot} except that the horizontal axis is time (IJD = days from 1/1/2000) rather than Galactic longitude.  Here we have used a $4 \sigma$ threshold rather than $5 \sigma$ as before since with these shorter accumulations the effects of systematic errors due to imperfect background subtraction are smaller.   We have individually examined all spectra in which line candidates are found.  The most frequent detections are lines of $\sim 60$  keV energy and $\sim 15-20$  keV width.  These are clearly residuals from the strong background line complex in the 54-67 keV range (see Fig.~\ref{bckgnd}).  The most significant detections are a pair of lines at 58 and 76 keV.  This pair appears 5 times in the plot.  After examining the data we found that they are produced by energetic electrons that leaked from the earth's magnetosphere and escaped our filter.  Our filter rejects data 8 hr before and 4 hr after perigee, but occasionally magnetospheric particles can appear outside of this window.  The detections are fluorescence lines activated by bremsstrahlung photons from the magnetospheric electrons.  The 58 keV line is the K$_{\alpha}$ line from tungsten (the material of the mask), and the 76 keV line is K$_{\alpha}$ line from bismuth (one of the main components of the shield).  Again, these detections serve to validate our procedures.   All other detections are just above the $4 \sigma$ threshold and are expected since the number of trials is large (6004 spectra x 2800 energy bins/spectrum).  From this analysis and that described in Section~\ref{smed_res} we conclude that there is no evidence for variable lines in the INTEGRAL/SPI data on time scales from a fraction of an hour to $\sim 1$ yr.  

\section{Discussion and Conclusions}

We have carried out a comprehensive search of the first year of the INTEGRAL/SPI data for gamma-ray lines and found no new lines.  Our search has spanned a broad range in line energy ($\sim40$ - 8000 keV), width (3 - 1000 keV) and variability time scale ($\sim 1000$ sec. to $\sim 1$ yr).  We have searched for both point-like and diffuse sources.  We used the INTEGRAL Bright Source Catalog as an input and searched the spectra of 143 known sources.  The diffuse source search covered the spatial range from $\sim 20 \degr$ (the approximate field-of-view of the spectrometer) up to the entire Galactic plane. This search would also have detected previously unknown point sources with line fluxes exceeding our detection threshold.  The sensitivities varied over a wide range (a few $\times 10^{-5}$ to a few $\times 10^{-2} \mbox{ cm}^{-2} \mbox{ s}^{-1} $) depending on line energy and width, spatial scale, and variability time scale.  

\begin{figure}
\plotone{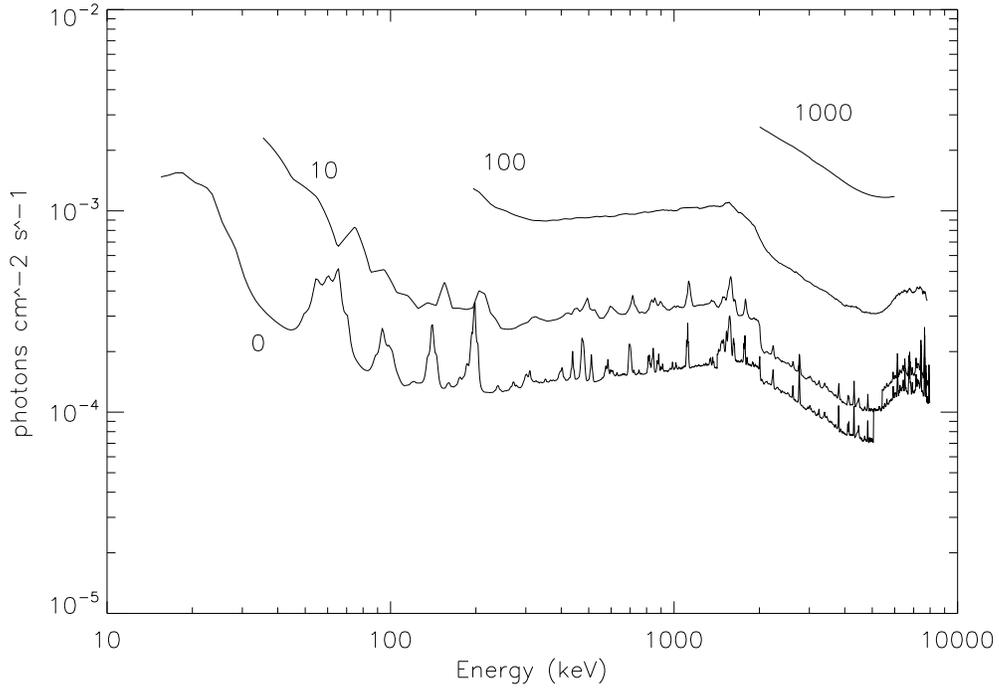}
\caption{Line search sensitivity ($3.5 \sigma$) for Crab.  Curves are labeled with assumed line width (keV FWHM). Values in the 1400-1600 keV interval should be ignored as this region was excluded from analysis due to the presence of instrumental noise.  The step at 2000 keV is due to the change in binning discussed in Section 3.3 and the accompanying reduction in the errors returned by the by SPIROS program. The steps at $\sim 4$ MeV correspond to the point at which the increasing (with energy) instrumental line width forces a change in the optimal template width for the line detection. }
\label{crab_sens}
\end{figure}

We discussed various predictions of gamma-ray lines in Section~\ref{grl_prod}.   In most cases the predictions were below the current INTEGRAL/SPI sensitivity thresholds.   For pulsars, \citet{rom96} has predicted a positron production rate of $\sim 10^{38} \mbox{ s}^{-1}$, which leads to a 511 keV flux of $\sim 2 \times 10^{-7} \mbox{ cm}^{-2} \mbox{ s}^{-1} $ at the earth.  For comparison, the INTEGRAL/SPI 511 keV sensitivities for two pulsars PSR B1509-38 and Vela (see Table~\ref{sens_pulsar}) are $1-2 \times 10^{-4} \mbox{ cm}^{-2} \mbox{ s}^{-1} $, well above the prediction.  For neutron stars, \citet{bil93}  have investigated the production of gamma-rays through accretion-induced surface reactions.  The strongest predicted line is that from neutron-proton recombination (2223 keV).  Neutrons are created by spallation of accreting helium and heavier nuclei.  The predicted flux from Sco X-1 is $\sim 10^{-6} \mbox{ cm}^{-2} \mbox{ s}^{-1} $ with fluxes as high as $\sim 10^{-5} \mbox{ cm}^{-2} \mbox{ s}^{-1} $  being possible.   The INTEGRAL/SPI  2223 keV sensitivity for Sco X-1 can be found by applying the scale factor in Table~\ref{sens_lmxb} to Fig.~\ref{crab_sens}.  The derived value is $\sim 10^{-4} \mbox{ cm}^{-2} \mbox{ s}^{-1} $,  about an order of magnitude larger than the most optimistic prediction.  The INTEGRAL Sco X-1 sensitivity can be expected to improve over the course of the mission since we have only analyzed the first year of data in which there were $\sim 7 \times 10^5$ sec. of useful data from Sco X-1.  

We mentioned in Section~\ref{ep_ann} reports from the Granat/SIGMA experiment of strong transient line-like features in the vicinity of 500 keV from 1E1740.7-2942  and Nova Musca.  The fluxes were $\sim1.2 \times 10^{-2} \mbox{ cm}^{-2} \mbox{ s}^{-1} $ and $6 \pm 3 \times  10^{-3} \mbox{ cm}^{-2} \mbox{ s}^{-1} $, respectively, lasting for $\sim 1$ day.  The line widths were $\sim 240$ keV and $\sim 23$ keV, respectively.  For a 1-day integration the INTEGRAL/SPI sensitivities at 500 keV are $\sim 10^{-3} \mbox{ cm}^{-2} \mbox{ s}^{-1}$ and $\sim 3 \times 10^{-4} \mbox{ cm}^{-2} \mbox{ s}^{-1}$ for line widths of 240 and 23 keV.   (See also \citet{win03b} for a discussion of the sensitivity of the INTEGRAL Galactic Plane scans to such events.)  Our search methods were sensitive to transient events of duration $\sim 1000$ sec up to $\sim 1$ yr so these events would have been easily detected by INTEGRAL.   Granat found two of them during its $\sim 10$ yr lifetime.  So far we have only analyzed 1 yr of INTEGRAL data.  However, INTEGRAL is more sensitive than Granat, so, if the events are real, INTEGRAL should have a reasonable chance of detecting one or more in its lifetime.   

\begin{figure}
\plotone{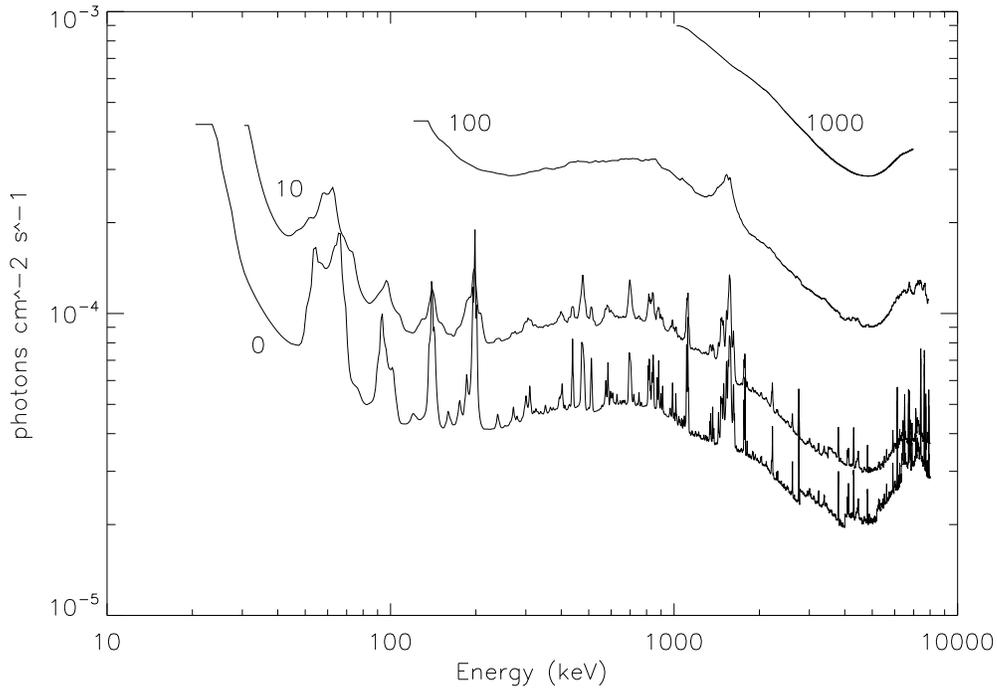}
\caption{Line search sensitivity ($3.5 \sigma$) for Galactic Center spectral accumulation ($\theta < 13 \degr$) and assumed $10 \degr$ gaussian spatial distribution (GC13,I).   Curves are labeled with assumed line width (keV FWHM).  Values in the 1400-1600 keV interval should be ignored as this region was excluded from analysis due to the presence of instrumental noise.}
\label{ls_sens}
\end{figure}

In Section~\ref{grl_prod} we discussed gamma-ray line production by novae.  The longer-lived isotopes $^{7}$Be ($t_{1/2} = 53.3$~d, $E_{\gamma} =$ 478~keV)  from CO novae and $^{22}$Na ($t_{1/2} = 2.6$~y, $E_{\gamma} =$~1275~keV) from ONe novae appeared to hold the most promise for an INTEGRAL/SPI detection.  The nova rate in our Galaxy is $\sim 40 \mbox{ yr}^{-1}$ \citep{cox00} roughly equally divided between CO and ONe novae.  We can therefore expect that the 478 keV line will result from the most recent $\sim 4$ CO novae and the 1275 keV line from the most recent $\sim 75$ ONe novae in our Galaxy.   \citet{gom98} predict fluxes of $\sim 2 \times 10^{-6} \mbox{ cm}^{-2} \mbox{ s}^{-1}$ for the 478 keV line and $\sim 5 \times 10^{-6} \mbox{ cm}^{-2} \mbox{ s}^{-1} $ for the 1275~keV line for a nova at a distance of 1~kpc.  Scaling to the Galactic Center (D~=~8~kpc) and multiplying by the expected number of nova in 1 decay time yield fluxes of $\sim 1.3 \times 10^{-7} \mbox{ cm}^{-2} \mbox{ s}^{-1}$ and $\sim 6 \times 10^{-6} \mbox{ cm}^{-2} \mbox{ s}^{-1}$ for the 478 and 1275 keV lines, respectively. These predictions should be compared to our detection limits for the scenario where a flat distribution of radius $30 \degr$ at the Galactic Center (GC13,II) is assumed.  From Table~\ref{sens_summ_lar} and Fig.~\ref{ls_sens} we deduce INTEGRAL/SPI sensitivities of $\sim 1.5 \times 10^{-4} \mbox{ cm}^{-2} \mbox{ s}^{-1}$ and $\sim 1.2 \times 10^{-4} \mbox{ cm}^{-2} \mbox{ s}^{-1}$ for the 478 and 1275 keV lines.  The 1275 keV prediction is a factor of $\sim 20$ below the current INTEGRAL/SPI threshold.  However, the predictions are uncertain, and the INTEGRAL sensitivity will grow over the course of the mission, so it is not out of the question that there will be an eventual detection.   

Perhaps the most promising scenario for the detection of new gamma-ray lines is from cosmic-ray excitation of ambient nuclei in the interstellar medium.  This process has been treated extensively by \citet{ram79}.  The principal uncertainties are the composition of the cosmic rays and the ambient medium and the energy spectrum and density of the cosmic rays.  The strongest line predicted by \citet{ram79} is from the de-excitation of $^{12}$C (4439 keV).  The largest predicted flux is $\sim 7 \times 10^{-5} \mbox{ cm}^{-2} \mbox{ s}^{-1} \mbox{ rad}^{-1}$ from the central region of our Galaxy.  The most directly comparable line search result is that in Table~\ref{sens_summ_lar} for the GC30, II case where the flux is assumed to be uniformly distributed over a circular region at the Galactic Center of radius $30 \degr$.   The INTEGRAL/SPI sensitivity at 4439 keV is $\sim 10^{-4} \mbox{ cm}^{-2} \mbox{ s}^{-1}$ and the value is roughly the same if calculated per radian.    This value is only slightly higher than the most optimistic prediction of  \citet{ram79}.  Over the course of the mission INTEGRAL will spend a sizable fraction of its time observing the central region of our Galaxy, and we can expect that the limits will significantly improve and provide a more stringent test of the model.   

\acknowledgements

INTEGRAL is a project of the European Space Agency to which NASA is a contributing partner.  We are grateful for valuable comments from P. Jean, J. Kn\"{o}dlseder, G. Skinner, G. Weidenspointner and the anonymous referee.

%%Tables***********************************
\clearpage

\begin{deluxetable}{lccl}
\tabletypesize{\scriptsize}
%\tablenum{1}
\tablecaption{Observed Astrophysical Gamma-Ray Lines}
\tablewidth{0pt}
\tablehead{\colhead{Line} & \colhead{Energy (keV)} & \colhead{Half-Life (yr)} &  \colhead{Source(s)}}
\startdata 
$e^{+}$ - $e^{-}$ annihilation & 511 & N/A & Galactic Center/Plane\\
&  & &1E1740.7-2942\\
& & & Nova Musca\\
$^{26}$Al & 1809 & $7.2 \times 10^{5} $ & Galactic Plane\\
$^{44}$Ti & 68, 78, 1157 & 63 & Cas A\\
& & & Vela Region\\
$^{60}$Fe & 59, 1173, 1333 & $1.52 \times 10^{6} $ & Galactic Plane\\
$^{56}$Co & 847, 1238 & 0.21 & SN1987A\\
$^{57}$Co & 122, 137 & 0.74 & SN1987A\\
$^{2}$H, $^{12}$C, $^{16}$O, ... & 2223, 4438, 6129, ... & prompt & Sun (solar flare)
\enddata
\label{obs_lines}
\end{deluxetable}

\begin{deluxetable}{ccc}
\tabletypesize{\scriptsize}
\tablecaption{Large-Scale Data Sets}
%\tablenum{1}
\tablewidth{0pt}
\tablehead{\colhead{Label} & \colhead{Description} & \colhead{Criterion}}
\startdata 
GC13 & (Galactic Center) - (Off-Center) & GC region: $<13 \degr$ from center\\
GC30 & (Galactic Center) - (Off-Center) & GC region: $<30 \degr$ from center\\
GP & (Galactic Plane) - (Off-Plane) & GP region: $\| \mbox{b} \| < 20 \degr$
\enddata
\label{ls_parms}
\end{deluxetable}

\begin{deluxetable}{cc}
\tabletypesize{\scriptsize}
\tablecaption{Diffuse Source Spatial Distributions}
%\tablenum{1}
\tablewidth{0pt}
\tablehead{\colhead{Label} & \colhead{Description}}
\startdata 
I & symmetric gaussian ($10 \degr$ FWHM) centered at GC\\
II & flat distribution (radius = $30 \degr$) centered at GC\\
III & flat in Galactic longitude, gaussian in Galactic latitude (FWHM = $ 5  \degr$)\\
IV & Dirbe $60 \micron$ \\
V & Egret 30 - 100 MeV	
\enddata
\label{spatial_distr}
\end{deluxetable}

\begin{deluxetable}{cccccccc}
\tabletypesize{\scriptsize}
\tablecaption{Large-Scale Sensitivities}
%\tablenum{1}
\tablewidth{0pt}
\tablehead{\colhead{Data Set} & \colhead{Spatial Distribution}& \colhead{Exposure (sec)} & \colhead{100 keV\tablenotemark{a}} & \colhead{500 keV\tablenotemark{a}}  & \colhead{1000 keV\tablenotemark{a}} & \colhead{5000 keV\tablenotemark{a}} & \colhead{Scale Factor\tablenotemark{b}}}
\startdata 
GC13 & I & $1.905 \times 10^{6}$ & $6.53 \times 10^{-5}$ & $4.87 \times 10^{-5}$ & $4.61 \times 10^{-5}$ & $2.07 \times 10^{-5}$ &   1.00\\
GC30 & II & $3.621 \times 10^{6}$ & $2.22 \times 10^{-4}$ & $1.50 \times 10^{-4}$ & $1.30 \times 10^{-4}$ & $4.80 \times 10^{-5}$ &   3.08\\
GP & III & $3.846 \times 10^{6}$ & $6.22 \times 10^{-4}$ & $4.41 \times 10^{-4}$ & $4.01 \times 10^{-4}$ & $1.64 \times 10^{-4}$ &  9.05\\
GP & IV & $3.846\times 10^{6}$ & $3.17 \times 10^{-4}$ & $2.26 \times 10^{-4}$ & $2.05 \times 10^{-4}$ & $8.37 \times 10^{-5}$ &  4.63\\
GP & V & $3.846 \times 10^{6}$ & $9.35 \times 10^{-4}$ & $6.43 \times 10^{-4}$ & $5.70 \times 10^{-4}$ & $2.20 \times 10^{-4}$ &  13.21
\enddata
\tablenotetext{a}{Narrow line sensitivity [total Galactic flux (photons cm$^{-2}$ sec$^{-1}$)]}
\tablenotetext{b}{Scale factor to apply to sensitivity plot of Fig. ~\ref{ls_sens}.}
\label{sens_summ_lar}
\end{deluxetable}

\begin{deluxetable}{cccccccc}
\tabletypesize{\scriptsize}
\tablecaption{Small-to-Medium-Scale Sensitivities}
%\tablenum{1}
\tablewidth{0pt}
\tablehead{\colhead{l} & \colhead{b} & \colhead{Exposure (sec)} & \colhead{100 keV\tablenotemark{a}} & \colhead{511 keV\tablenotemark{a}}  & \colhead{1000 keV\tablenotemark{a}} & \colhead{5000 keV\tablenotemark{a}} & \colhead{Scale Factor\tablenotemark{b}}}
\startdata 
 180.0 &  -54.0 & $2.560 \times 10^{5}$ & $1.86 \times 10^{-4}$ & $1.85 \times 10^{-4}$ & $1.37 \times 10^{-4}$ & $6.30 \times 10^{-5}$ &   2.16\\
-180.0 &  -18.0 & $1.309 \times 10^{5}$ & $3.53 \times 10^{-4}$ & $3.45 \times 10^{-4}$ & $2.54 \times 10^{-4}$ & $1.17 \times 10^{-4}$ &   4.02\\
 180.0 &    0.0 & $4.191 \times 10^{5}$ & $1.69 \times 10^{-4}$ & $1.66 \times 10^{-4}$ & $1.21 \times 10^{-4}$ & $5.56 \times 10^{-5}$ &   1.93\\
-180.0 &   72.0 & $3.656 \times 10^{4}$ & $5.44 \times 10^{-4}$ & $5.45 \times 10^{-4}$ & $3.99 \times 10^{-4}$ & $1.83 \times 10^{-4}$ &   6.36\\
-162.0 &    0.0 & $3.666 \times 10^{5}$ & $1.80 \times 10^{-4}$ & $1.77 \times 10^{-4}$ & $1.30 \times 10^{-4}$ & $5.93 \times 10^{-5}$ &   2.07\\
-158.0 &  -18.0 & $4.880 \times 10^{4}$ & $4.39 \times 10^{-4}$ & $4.30 \times 10^{-4}$ & $3.17 \times 10^{-4}$ & $1.46 \times 10^{-4}$ &   5.02\\
-144.0 &    0.0 & $2.250 \times 10^{4}$ & $6.19 \times 10^{-4}$ & $6.14 \times 10^{-4}$ & $4.53 \times 10^{-4}$ & $2.09 \times 10^{-4}$ &   7.17\\
-126.0 &    0.0 & $2.507 \times 10^{4}$ & $5.79 \times 10^{-4}$ & $5.77 \times 10^{-4}$ & $4.26 \times 10^{-4}$ & $1.97 \times 10^{-4}$ &   6.74\\
-108.0 &    0.0 & $6.608 \times 10^{4}$ & $3.78 \times 10^{-4}$ & $3.77 \times 10^{-4}$ & $2.81 \times 10^{-4}$ & $1.30 \times 10^{-4}$ &   4.40\\
 -90.0 &  -36.0 & $3.837 \times 10^{5}$ & $1.50 \times 10^{-4}$ & $1.49 \times 10^{-4}$ & $1.11 \times 10^{-4}$ & $5.11 \times 10^{-5}$ &   1.74\\
 -90.0 &    0.0 & $1.439 \times 10^{5}$ & $2.40 \times 10^{-4}$ & $2.40 \times 10^{-4}$ & $1.79 \times 10^{-4}$ & $8.29 \times 10^{-5}$ &   2.80\\
 -90.0 &   54.0 & $1.281 \times 10^{5}$ & $2.58 \times 10^{-4}$ & $2.58 \times 10^{-4}$ & $1.92 \times 10^{-4}$ & $8.87 \times 10^{-5}$ &   3.02\\
 -90.0 &   72.0 & $1.933 \times 10^{5}$ & $2.11 \times 10^{-4}$ & $2.10 \times 10^{-4}$ & $1.56 \times 10^{-4}$ & $7.20 \times 10^{-5}$ &   2.46\\
 -72.0 &    0.0 & $5.436 \times 10^{4}$ & $3.78 \times 10^{-4}$ & $3.78 \times 10^{-4}$ & $2.82 \times 10^{-4}$ & $1.30 \times 10^{-4}$ &   4.41\\
 -67.0 &  -18.0 & $4.052 \times 10^{4}$ & $4.83 \times 10^{-4}$ & $4.81 \times 10^{-4}$ & $3.56 \times 10^{-4}$ & $1.63 \times 10^{-4}$ &   5.62\\
 -60.0 &  -36.0 & $9.122 \times 10^{4}$ & $3.16 \times 10^{-4}$ & $3.15 \times 10^{-4}$ & $2.33 \times 10^{-4}$ & $1.07 \times 10^{-4}$ &   3.68\\
 -60.0 &   36.0 & $3.583 \times 10^{4}$ & $4.93 \times 10^{-4}$ & $4.92 \times 10^{-4}$ & $3.62 \times 10^{-4}$ & $1.66 \times 10^{-4}$ &   5.74\\
 -54.0 &    0.0 & $4.535 \times 10^{5}$ & $1.71 \times 10^{-4}$ & $1.70 \times 10^{-4}$ & $1.27 \times 10^{-4}$ & $5.86 \times 10^{-5}$ &   1.99\\
 -45.0 &   18.0 & $2.543 \times 10^{5}$ & $1.85 \times 10^{-4}$ & $1.84 \times 10^{-4}$ & $1.35 \times 10^{-4}$ & $6.22 \times 10^{-5}$ &   2.14\\
 -45.0 &   54.0 & $5.891 \times 10^{4}$ & $3.85 \times 10^{-4}$ & $3.85 \times 10^{-4}$ & $2.86 \times 10^{-4}$ & $1.32 \times 10^{-4}$ &   4.50\\
 -36.0 &    0.0 & $5.314 \times 10^{5}$ & $1.32 \times 10^{-4}$ & $1.31 \times 10^{-4}$ & $9.64 \times 10^{-5}$ & $4.43 \times 10^{-5}$ &   1.53\\
 -30.0 &   36.0 & $1.104 \times 10^{5}$ & $2.82 \times 10^{-4}$ & $2.81 \times 10^{-4}$ & $2.07 \times 10^{-4}$ & $9.51 \times 10^{-5}$ &   3.28\\
 -23.0 &   18.0 & $3.586 \times 10^{5}$ & $1.53 \times 10^{-4}$ & $1.53 \times 10^{-4}$ & $1.12 \times 10^{-4}$ & $5.16 \times 10^{-5}$ &   1.79\\
 -22.0 &  -18.0 & $1.914 \times 10^{5}$ & $2.12 \times 10^{-4}$ & $2.12 \times 10^{-4}$ & $1.55 \times 10^{-4}$ & $7.09 \times 10^{-5}$ &   2.47\\
 -18.0 &    0.0 & $5.644 \times 10^{5}$ & $1.21 \times 10^{-4}$ & $1.20 \times 10^{-4}$ & $8.84 \times 10^{-5}$ & $4.06 \times 10^{-5}$ &   1.40\\
   0.0 &  -18.0 & $2.853 \times 10^{5}$ & $1.74 \times 10^{-4}$ & $1.73 \times 10^{-4}$ & $1.27 \times 10^{-4}$ & $5.81 \times 10^{-5}$ &   2.02\\
   0.0 &    0.0 & $1.165 \times 10^{6}$ & $9.57 \times 10^{-5}$ & $9.51 \times 10^{-5}$ & $6.97 \times 10^{-5}$ & $3.20 \times 10^{-5}$ &   1.11\\
   0.0 &   18.0 & $4.576 \times 10^{5}$ & $1.39 \times 10^{-4}$ & $1.38 \times 10^{-4}$ & $1.01 \times 10^{-4}$ & $4.66 \times 10^{-5}$ &   1.61\\
   0.0 &   36.0 & $1.371 \times 10^{5}$ & $2.59 \times 10^{-4}$ & $2.58 \times 10^{-4}$ & $1.91 \times 10^{-4}$ & $8.79 \times 10^{-5}$ &   3.02\\
   0.0 &   72.0 & $2.891 \times 10^{4}$ & $5.42 \times 10^{-4}$ & $5.42 \times 10^{-4}$ & $4.01 \times 10^{-4}$ & $1.85 \times 10^{-4}$ &   6.33\\
  18.0 &    0.0 & $4.222 \times 10^{5}$ & $1.53 \times 10^{-4}$ & $1.53 \times 10^{-4}$ & $1.12 \times 10^{-4}$ & $5.11 \times 10^{-5}$ &   1.79\\
  23.0 &  -18.0 & $1.463 \times 10^{5}$ & $2.44 \times 10^{-4}$ & $2.45 \times 10^{-4}$ & $1.78 \times 10^{-4}$ & $8.12 \times 10^{-5}$ &   2.86\\
  23.0 &   18.0 & $1.725 \times 10^{5}$ & $2.29 \times 10^{-4}$ & $2.30 \times 10^{-4}$ & $1.67 \times 10^{-4}$ & $7.62 \times 10^{-5}$ &   2.68\\
  30.0 &  -36.0 & $2.759 \times 10^{4}$ & $9.86 \times 10^{-4}$ & $9.79 \times 10^{-4}$ & $7.25 \times 10^{-4}$ & $3.37 \times 10^{-4}$ &  11.43\\
  36.0 &    0.0 & $8.999 \times 10^{5}$ & $1.00 \times 10^{-4}$ & $9.94 \times 10^{-5}$ & $7.33 \times 10^{-5}$ & $3.38 \times 10^{-5}$ &   1.16\\
  37.0 &   90.0 & $3.360 \times 10^{5}$ & $1.58 \times 10^{-4}$ & $1.58 \times 10^{-4}$ & $1.17 \times 10^{-4}$ & $5.40 \times 10^{-5}$ &   1.85\\
  45.0 &  -54.0 & $5.093 \times 10^{4}$ & $4.02 \times 10^{-4}$ & $4.03 \times 10^{-4}$ & $2.98 \times 10^{-4}$ & $1.38 \times 10^{-4}$ &   4.70\\
  45.0 &  -18.0 & $6.642 \times 10^{4}$ & $3.56 \times 10^{-4}$ & $3.52 \times 10^{-4}$ & $2.61 \times 10^{-4}$ & $1.21 \times 10^{-4}$ &   4.11\\
  54.0 &    0.0 & $3.763 \times 10^{5}$ & $1.51 \times 10^{-4}$ & $1.50 \times 10^{-4}$ & $1.11 \times 10^{-4}$ & $5.11 \times 10^{-5}$ &   1.75\\
  68.0 &   18.0 & $1.551 \times 10^{5}$ & $2.43 \times 10^{-4}$ & $2.35 \times 10^{-4}$ & $1.76 \times 10^{-4}$ & $8.18 \times 10^{-5}$ &   2.74\\
  72.0 &    0.0 & $2.961 \times 10^{5}$ & $2.11 \times 10^{-4}$ & $2.07 \times 10^{-4}$ & $1.54 \times 10^{-4}$ & $7.16 \times 10^{-5}$ &   2.41\\
  90.0 &  -54.0 & $1.589 \times 10^{4}$ & $1.23 \times 10^{-3}$ & $1.23 \times 10^{-3}$ & $9.24 \times 10^{-4}$ & $4.28 \times 10^{-4}$ &  14.32\\
  90.0 &  -36.0 & $1.589 \times 10^{4}$ & $1.23 \times 10^{-3}$ & $1.23 \times 10^{-3}$ & $9.24 \times 10^{-4}$ & $4.28 \times 10^{-4}$ &  14.32\\
  90.0 &  -18.0 & $1.979 \times 10^{4}$ & $6.50 \times 10^{-4}$ & $6.39 \times 10^{-4}$ & $4.77 \times 10^{-4}$ & $2.21 \times 10^{-4}$ &   7.46\\
  90.0 &    0.0 & $2.472 \times 10^{5}$ & $1.86 \times 10^{-4}$ & $1.82 \times 10^{-4}$ & $1.35 \times 10^{-4}$ & $6.30 \times 10^{-5}$ &   2.12\\
  90.0 &   54.0 & $4.553 \times 10^{4}$ & $4.21 \times 10^{-4}$ & $4.21 \times 10^{-4}$ & $3.11 \times 10^{-4}$ & $1.43 \times 10^{-4}$ &   4.92\\
  90.0 &   72.0 & $1.462 \times 10^{5}$ & $2.56 \times 10^{-4}$ & $2.57 \times 10^{-4}$ & $1.89 \times 10^{-4}$ & $8.71 \times 10^{-5}$ &   2.99\\
 108.0 &    0.0 & $1.783 \times 10^{5}$ & $2.16 \times 10^{-4}$ & $2.16 \times 10^{-4}$ & $1.59 \times 10^{-4}$ & $7.35 \times 10^{-5}$ &   2.52\\
 126.0 &    0.0 & $1.257 \times 10^{5}$ & $2.61 \times 10^{-4}$ & $2.60 \times 10^{-4}$ & $1.92 \times 10^{-4}$ & $8.84 \times 10^{-5}$ &   3.04\\
 135.0 &   54.0 & $6.624 \times 10^{4}$ & $3.50 \times 10^{-4}$ & $3.52 \times 10^{-4}$ & $2.59 \times 10^{-4}$ & $1.19 \times 10^{-4}$ &   4.10\\
 144.0 &    0.0 & $1.746 \times 10^{4}$ & $7.06 \times 10^{-4}$ & $7.00 \times 10^{-4}$ & $5.16 \times 10^{-4}$ & $2.38 \times 10^{-4}$ &   8.17\\
 162.0 &    0.0 & $7.835 \times 10^{4}$ & $4.97 \times 10^{-4}$ & $4.92 \times 10^{-4}$ & $3.63 \times 10^{-4}$ & $1.67 \times 10^{-4}$ &   5.74
\enddata
\tablenotetext{a}{Narrow line sensitivity (photons cm$^{-2}$ sec$^{-1}$)}
\tablenotetext{b}{Scale factor to apply to sensitivity plot of Fig. ~\ref{ls_sens}.}
\label{sens_summ_smed}
\end{deluxetable}

\begin{deluxetable}{lccccccc}
\tabletypesize{\scriptsize}
\tablecaption{Point Source Sensitivities - LMXB}
%\tablenum{6a}
\tablewidth{0pt}
\tablehead{\colhead{Source} & \colhead{Exposure (sec)} & \colhead{50 keV\tablenotemark{a}} & \colhead{100 keV\tablenotemark{a}}  & \colhead{511 keV\tablenotemark{a}} & \colhead{1000 keV\tablenotemark{a}} & \colhead{Scale Factor\tablenotemark{b}}}
\startdata 
1A 1742-294 & $4.270 \times 10^{6}$ & $1.32 \times 10^{-3}$ & $1.66 \times 10^{-3}$ & $8.36 \times 10^{-4}$ & $6.55 \times 10^{-4}$ &   5.33\\
1E 1740 & $4.282 \times 10^{6}$ & $9.56 \times 10^{-4}$ & $1.18 \times 10^{-3}$ & $6.00 \times 10^{-4}$ & $4.62 \times 10^{-4}$ &   3.76\\
3A 1728-169 & $4.426 \times 10^{6}$ & $1.69 \times 10^{-4}$ & $2.11 \times 10^{-4}$ & $1.09 \times 10^{-4}$ & $8.44 \times 10^{-5}$ &   0.69\\
3A 1822-371 & $4.346 \times 10^{6}$ & $1.91 \times 10^{-4}$ & $2.29 \times 10^{-4}$ & $1.19 \times 10^{-4}$ & $9.33 \times 10^{-5}$ &   0.76\\
4U 1626-67 & $8.267 \times 10^{5}$ & $4.66 \times 10^{-4}$ & $5.47 \times 10^{-4}$ & $2.90 \times 10^{-4}$ & $2.54 \times 10^{-4}$ &   2.07\\
4U 1730-335 & $4.181 \times 10^{6}$ & $3.17 \times 10^{-4}$ & $3.47 \times 10^{-4}$ & $2.07 \times 10^{-4}$ & $1.61 \times 10^{-4}$ &   1.31\\
4U 1735-444 & $4.236 \times 10^{6}$ & $2.05 \times 10^{-4}$ & $2.44 \times 10^{-4}$ & $1.21 \times 10^{-4}$ & $1.02 \times 10^{-4}$ &   0.83\\
4U 1916-053 & $1.552 \times 10^{6}$ & $2.53 \times 10^{-4}$ & $2.98 \times 10^{-4}$ & $1.61 \times 10^{-4}$ & $1.47 \times 10^{-4}$ &   1.19\\
Aql X-1 & $1.073 \times 10^{6}$ & $2.58 \times 10^{-4}$ & $3.47 \times 10^{-4}$ & $1.78 \times 10^{-4}$ & $1.64 \times 10^{-4}$ &   1.34\\
Cir X-1 & $1.853 \times 10^{6}$ & $2.60 \times 10^{-4}$ & $3.44 \times 10^{-4}$ & $1.71 \times 10^{-4}$ & $1.39 \times 10^{-4}$ &   1.14\\
Cyg X-2 & $5.252 \times 10^{5}$ & $4.32 \times 10^{-4}$ & $5.18 \times 10^{-4}$ & $2.50 \times 10^{-4}$ & $2.06 \times 10^{-4}$ &   1.68\\
EXO 0748-676 & $7.796 \times 10^{5}$ & $2.29 \times 10^{-4}$ & $2.92 \times 10^{-4}$ & $1.41 \times 10^{-4}$ & $1.27 \times 10^{-4}$ &   1.04\\
GRO J1655-40 & $4.052 \times 10^{6}$ & $1.84 \times 10^{-4}$ & $2.37 \times 10^{-4}$ & $1.21 \times 10^{-4}$ & $1.01 \times 10^{-4}$ &   0.82\\
GRS 1739-278 & $4.196 \times 10^{6}$ & $2.11 \times 10^{-4}$ & $2.57 \times 10^{-4}$ & $1.33 \times 10^{-4}$ & $1.03 \times 10^{-4}$ &   0.84\\
GX 1+4 & $4.196 \times 10^{6}$ & $1.41 \times 10^{-4}$ & $1.82 \times 10^{-4}$ & $9.49 \times 10^{-5}$ & $7.36 \times 10^{-5}$ &   0.60\\
GX 13+1 & $4.060 \times 10^{6}$ & $1.79 \times 10^{-4}$ & $2.40 \times 10^{-4}$ & $1.25 \times 10^{-4}$ & $9.93 \times 10^{-5}$ &   0.81\\
GX 17+2 & $3.732 \times 10^{6}$ & $2.12 \times 10^{-4}$ & $2.59 \times 10^{-4}$ & $1.35 \times 10^{-4}$ & $1.08 \times 10^{-4}$ &   0.88\\
GX 3+1 & $4.196 \times 10^{6}$ & $1.59 \times 10^{-4}$ & $1.94 \times 10^{-4}$ & $1.01 \times 10^{-4}$ & $7.86 \times 10^{-5}$ &   0.64\\
GX 349+2 & $4.320 \times 10^{6}$ & $1.79 \times 10^{-4}$ & $2.28 \times 10^{-4}$ & $1.16 \times 10^{-4}$ & $9.28 \times 10^{-5}$ &   0.76\\
GX 354-0 & $4.354 \times 10^{6}$ & $3.15 \times 10^{-4}$ & $3.95 \times 10^{-4}$ & $2.09 \times 10^{-4}$ & $1.62 \times 10^{-4}$ &   1.32\\
GX 5-1 & $4.318 \times 10^{6}$ & $1.43 \times 10^{-4}$ & $1.90 \times 10^{-4}$ & $9.83 \times 10^{-5}$ & $7.62 \times 10^{-5}$ &   0.62\\
GX 9+1 & $4.116 \times 10^{6}$ & $2.31 \times 10^{-4}$ & $2.87 \times 10^{-4}$ & $1.49 \times 10^{-4}$ & $1.16 \times 10^{-4}$ &   0.94\\
Ginga 1826-24 & $4.116 \times 10^{6}$ & $1.65 \times 10^{-4}$ & $2.10 \times 10^{-4}$ & $1.09 \times 10^{-4}$ & $8.47 \times 10^{-5}$ &   0.69\\
H 0614+091 & $6.176 \times 10^{5}$ & $4.86 \times 10^{-4}$ & $5.35 \times 10^{-4}$ & $2.54 \times 10^{-4}$ & $1.84 \times 10^{-4}$ &   1.50\\
H 1608-522 & $1.708 \times 10^{6}$ & $2.23 \times 10^{-4}$ & $2.95 \times 10^{-4}$ & $1.47 \times 10^{-4}$ & $1.36 \times 10^{-4}$ &   1.11\\
H 1636-536 & $1.559 \times 10^{6}$ & $2.32 \times 10^{-4}$ & $2.85 \times 10^{-4}$ & $1.54 \times 10^{-4}$ & $1.39 \times 10^{-4}$ &   1.13\\
H 1702-429 & $4.062 \times 10^{6}$ & $2.55 \times 10^{-4}$ & $3.10 \times 10^{-4}$ & $1.63 \times 10^{-4}$ & $1.40 \times 10^{-4}$ &   1.14\\
H 1705-250 & $4.761 \times 10^{6}$ & $1.55 \times 10^{-4}$ & $1.90 \times 10^{-4}$ & $9.84 \times 10^{-5}$ & $7.62 \times 10^{-5}$ &   0.62\\
H 1705-440 & $3.859 \times 10^{6}$ & $2.43 \times 10^{-4}$ & $3.09 \times 10^{-4}$ & $1.56 \times 10^{-4}$ & $1.37 \times 10^{-4}$ &   1.11\\
H 1820-303 & $4.141 \times 10^{6}$ & $1.55 \times 10^{-4}$ & $1.96 \times 10^{-4}$ & $1.00 \times 10^{-4}$ & $7.66 \times 10^{-5}$ &   0.62\\
IGR J16358-4726 & $2.066 \times 10^{6}$ & $3.87 \times 10^{-4}$ & $4.79 \times 10^{-4}$ & $2.46 \times 10^{-4}$ & $2.24 \times 10^{-4}$ &   1.83\\
IGR J16418-4532 & $1.915 \times 10^{6}$ & $2.73 \times 10^{-4}$ & $3.50 \times 10^{-4}$ & $1.77 \times 10^{-4}$ & $1.61 \times 10^{-4}$ &   1.31\\
IGR J17464-3213 & $4.455 \times 10^{6}$ & $1.47 \times 10^{-4}$ & $1.96 \times 10^{-4}$ & $1.02 \times 10^{-4}$ & $8.07 \times 10^{-5}$ &   0.66\\
KS 1741-293 & $4.270 \times 10^{6}$ & $9.17 \times 10^{-3}$ & $1.14 \times 10^{-2}$ & $5.70 \times 10^{-3}$ & $4.52 \times 10^{-3}$ &  36.83\\
Sco X-1 & $6.764 \times 10^{5}$ & $2.78 \times 10^{-4}$ & $3.79 \times 10^{-4}$ & $2.00 \times 10^{-4}$ & $1.47 \times 10^{-4}$ &   1.20\\
Ser X-1 & $1.413 \times 10^{6}$ & $2.32 \times 10^{-4}$ & $3.01 \times 10^{-4}$ & $1.52 \times 10^{-4}$ & $1.39 \times 10^{-4}$ &   1.14\\
XTE J1748-288 & $4.196 \times 10^{6}$ & $2.78 \times 10^{-4}$ & $3.33 \times 10^{-4}$ & $1.73 \times 10^{-4}$ & $1.32 \times 10^{-4}$ &   1.08\\
XTE J1807-294 & $4.187 \times 10^{6}$ & $1.44 \times 10^{-4}$ & $1.87 \times 10^{-4}$ & $9.65 \times 10^{-5}$ & $7.43 \times 10^{-5}$ &   0.61\\
\\
LMXB Average & $1.219 \times 10^{8}$ & $3.47 \times 10^{-5}$ & $4.36 \times 10^{-5}$ & $2.26 \times 10^{-5}$ & $1.82 \times 10^{-5}$ &   0.15\\
\enddata
\tablenotetext{a}{Narrow line sensitivity (photons cm$^{-2}$ sec$^{-1}$)}
\tablenotetext{b}{Scale factor to apply to sensitivity plot of Fig.~\ref{crab_sens}.}
\label{sens_lmxb}
\end{deluxetable}

\begin{deluxetable}{lccccccc}
\tabletypesize{\scriptsize}
\tablecaption{Point Source Sensitivities - HMXB}
%\tablenum{6a}
\tablewidth{0pt}
\tablehead{\colhead{Source} & \colhead{Exposure (sec)} & \colhead{50 keV\tablenotemark{a}} & \colhead{100 keV\tablenotemark{a}}  & \colhead{511 keV\tablenotemark{a}} & \colhead{1000 keV\tablenotemark{a}} & \colhead{Scale Factor\tablenotemark{b}}}
\startdata 
1E 1145 & $8.413 \times 10^{5}$ & $3.28 \times 10^{-4}$ & $4.42 \times 10^{-4}$ & $2.21 \times 10^{-4}$ & $1.75 \times 10^{-4}$ &   1.43\\
3A 0114+650 & $3.109 \times 10^{5}$ & $8.06 \times 10^{-4}$ & $1.05 \times 10^{-3}$ & $5.26 \times 10^{-4}$ & $4.43 \times 10^{-4}$ &   3.61\\
3A 2206+543 & $4.375 \times 10^{5}$ & $5.14 \times 10^{-4}$ & $6.53 \times 10^{-4}$ & $3.32 \times 10^{-4}$ & $2.85 \times 10^{-4}$ &   2.32\\
4U 1036-56 & $2.016 \times 10^{5}$ & $7.82 \times 10^{-4}$ & $9.98 \times 10^{-4}$ & $4.97 \times 10^{-4}$ & $4.35 \times 10^{-4}$ &   3.54\\
4U 1700-377 & $4.332 \times 10^{6}$ & $1.78 \times 10^{-4}$ & $2.33 \times 10^{-4}$ & $1.20 \times 10^{-4}$ & $9.72 \times 10^{-5}$ &   0.79\\
4U 1901+03 & $1.494 \times 10^{6}$ & $2.22 \times 10^{-4}$ & $2.91 \times 10^{-4}$ & $1.48 \times 10^{-4}$ & $1.35 \times 10^{-4}$ &   1.10\\
Cen X-3 & $3.955 \times 10^{5}$ & $3.81 \times 10^{-4}$ & $5.06 \times 10^{-4}$ & $2.59 \times 10^{-4}$ & $2.30 \times 10^{-4}$ &   1.87\\
Cyg X-1 & $1.417 \times 10^{6}$ & $1.75 \times 10^{-4}$ & $2.21 \times 10^{-4}$ & $1.11 \times 10^{-4}$ & $9.34 \times 10^{-5}$ &   0.76\\
GX 301-2 & $8.190 \times 10^{5}$ & $2.93 \times 10^{-4}$ & $3.84 \times 10^{-4}$ & $1.93 \times 10^{-4}$ & $1.46 \times 10^{-4}$ &   1.19\\
Ginga 1843+009 & $1.494 \times 10^{6}$ & $2.16 \times 10^{-4}$ & $2.77 \times 10^{-4}$ & $1.43 \times 10^{-4}$ & $1.30 \times 10^{-4}$ &   1.06\\
H 0115+634 & $3.109 \times 10^{5}$ & $7.98 \times 10^{-4}$ & $1.03 \times 10^{-3}$ & $5.14 \times 10^{-4}$ & $4.33 \times 10^{-4}$ &   3.53\\
H 1538-522 & $1.723 \times 10^{6}$ & $2.39 \times 10^{-4}$ & $3.08 \times 10^{-4}$ & $1.54 \times 10^{-4}$ & $1.37 \times 10^{-4}$ &   1.12\\
H 1907+097 & $1.123 \times 10^{6}$ & $6.69 \times 10^{-4}$ & $8.83 \times 10^{-4}$ & $4.47 \times 10^{-4}$ & $4.14 \times 10^{-4}$ &   3.37\\
IGR J16318-4848 & $1.759 \times 10^{6}$ & $3.58 \times 10^{-4}$ & $4.72 \times 10^{-4}$ & $2.36 \times 10^{-4}$ & $2.16 \times 10^{-4}$ &   1.76\\
IGR J16320-4751 & $1.980 \times 10^{6}$ & $4.81 \times 10^{-4}$ & $6.21 \times 10^{-4}$ & $3.10 \times 10^{-4}$ & $2.82 \times 10^{-4}$ &   2.29\\
IGR J17391-3021 & $4.463 \times 10^{6}$ & $2.28 \times 10^{-4}$ & $2.83 \times 10^{-4}$ & $1.46 \times 10^{-4}$ & $1.13 \times 10^{-4}$ &   0.92\\
KS 1947+300 & $1.188 \times 10^{6}$ & $2.16 \times 10^{-4}$ & $2.74 \times 10^{-4}$ & $1.38 \times 10^{-4}$ & $1.14 \times 10^{-4}$ &   0.93\\
LMC X-4 & $7.796 \times 10^{5}$ & $2.27 \times 10^{-4}$ & $3.07 \times 10^{-4}$ & $1.56 \times 10^{-4}$ & $1.40 \times 10^{-4}$ &   1.14\\
OAO 1657-415 & $4.052 \times 10^{6}$ & $2.05 \times 10^{-4}$ & $2.64 \times 10^{-4}$ & $1.32 \times 10^{-4}$ & $1.15 \times 10^{-4}$ &   0.94\\
RX J0053 & $1.077 \times 10^{5}$ & $1.29 \times 10^{-3}$ & $1.53 \times 10^{-3}$ & $7.80 \times 10^{-4}$ & $5.51 \times 10^{-4}$ &   4.49\\
SAX J2103 & $1.448 \times 10^{6}$ & $2.48 \times 10^{-4}$ & $2.96 \times 10^{-4}$ & $1.42 \times 10^{-4}$ & $1.19 \times 10^{-4}$ &   0.97\\
SMC X-1 & $1.077 \times 10^{5}$ & $1.04 \times 10^{-3}$ & $1.36 \times 10^{-3}$ & $6.94 \times 10^{-4}$ & $5.01 \times 10^{-4}$ &   4.09\\
Vela X-1 & $1.043 \times 10^{6}$ & $1.99 \times 10^{-4}$ & $2.63 \times 10^{-4}$ & $1.37 \times 10^{-4}$ & $1.08 \times 10^{-4}$ &   0.88\\
XTE J1855-026 & $1.593 \times 10^{6}$ & $2.39 \times 10^{-4}$ & $3.09 \times 10^{-4}$ & $1.54 \times 10^{-4}$ & $1.42 \times 10^{-4}$ &   1.16\\
XTE J1908+094 & $4.392 \times 10^{5}$ & $1.03 \times 10^{-3}$ & $1.32 \times 10^{-3}$ & $6.71 \times 10^{-4}$ & $6.25 \times 10^{-4}$ &   5.10\\
\\
HMXB Average & $3.386 \times 10^{7}$ & $5.59 \times 10^{-5}$ & $7.21 \times 10^{-5}$ & $3.64 \times 10^{-5}$ & $3.08 \times 10^{-5}$ &   0.25\\
\enddata
\tablenotetext{a}{Narrow line sensitivity [(photons cm$^{-2}$ sec$^{-1}$)}
\tablenotetext{b}{Scale factor to apply to sensitivity plot of Fig.~\ref{crab_sens}.}
\label{sens_hmxb}
\end{deluxetable}

\begin{deluxetable}{lccccccc}
\tabletypesize{\scriptsize}
\tablecaption{Point Source Sensitivities - Pulsars}
%\tablenum{6a}
\tablewidth{0pt}
\tablehead{\colhead{Source} & \colhead{Exposure (sec)} & \colhead{50 keV\tablenotemark{a}} & \colhead{100 keV\tablenotemark{a}}  & \colhead{511 keV\tablenotemark{a}} & \colhead{1000 keV\tablenotemark{a}} & \colhead{Scale Factor\tablenotemark{b}}}
\startdata 
PSR B1509-58 & $1.591 \times 10^{6}$ & $2.63 \times 10^{-4}$ & $3.43 \times 10^{-4}$ & $1.75 \times 10^{-4}$ & $1.39 \times 10^{-4}$ &   1.13\\
Vela Pulsar & $1.043 \times 10^{6}$ & $1.84 \times 10^{-4}$ & $2.46 \times 10^{-4}$ & $1.27 \times 10^{-4}$ & $1.01 \times 10^{-4}$ &   0.82\\
\\
Pulsar Average & $2.634 \times 10^{6}$ & $1.51 \times 10^{-4}$ & $2.00 \times 10^{-4}$ & $1.03 \times 10^{-4}$ & $8.14 \times 10^{-5}$ &   0.66\\
\enddata
\tablenotetext{a}{Narrow line sensitivity (photons cm$^{-2}$ sec$^{-1}$)}
\tablenotetext{b}{Scale factor to apply to sensitivity plot of Fig.~\ref{crab_sens}.}
\label{sens_pulsar}
\end{deluxetable}

\begin{deluxetable}{lccccccc}
\tabletypesize{\scriptsize}
\tablecaption{Point Source Sensitivities - Quasars}
%\tablenum{6a}
\tablewidth{0pt}
\tablehead{\colhead{Source} & \colhead{Exposure (sec)} & \colhead{50 keV\tablenotemark{a}} & \colhead{100 keV\tablenotemark{a}}  & \colhead{511 keV\tablenotemark{a}} & \colhead{1000 keV\tablenotemark{a}} & \colhead{Scale Factor\tablenotemark{b}}}
\startdata 
3C 279 & $3.672 \times 10^{5}$ & $3.31 \times 10^{-4}$ & $4.38 \times 10^{-4}$ & $2.21 \times 10^{-4}$ & $1.67 \times 10^{-4}$ &   1.36\\
PKS 0637-752 & $8.873 \times 10^{5}$ & $2.31 \times 10^{-4}$ & $3.12 \times 10^{-4}$ & $1.58 \times 10^{-4}$ & $1.40 \times 10^{-4}$ &   1.14\\
PKS 1622-297 & $3.838 \times 10^{6}$ & $2.51 \times 10^{-4}$ & $2.96 \times 10^{-4}$ & $1.47 \times 10^{-4}$ & $1.18 \times 10^{-4}$ &   0.96\\
PKS 1830-211 & $4.165 \times 10^{6}$ & $1.83 \times 10^{-4}$ & $2.25 \times 10^{-4}$ & $1.19 \times 10^{-4}$ & $9.32 \times 10^{-5}$ &   0.76\\\enddata
\tablenotetext{a}{Narrow line sensitivity (photons cm$^{-2}$ sec$^{-1}$)}
\tablenotetext{b}{Scale factor to apply to sensitivity plot of Fig.~\ref{crab_sens}.}
\label{sens_quasar}
\end{deluxetable}

\begin{deluxetable}{lccccccc}
\tabletypesize{\scriptsize}
\tablecaption{Point Source Sensitivities - Supernova Remnants}
%\tablenum{6a}
\tablewidth{0pt}
\tablehead{\colhead{Source} & \colhead{Exposure (sec)} & \colhead{50 keV\tablenotemark{a}} & \colhead{100 keV\tablenotemark{a}}  & \colhead{511 keV\tablenotemark{a}} & \colhead{1000 keV\tablenotemark{a}} & \colhead{Scale Factor\tablenotemark{b}}}
\startdata 
Cas A & $3.659 \times 10^{5}$ & $5.02 \times 10^{-4}$ & $6.63 \times 10^{-4}$ & $3.32 \times 10^{-4}$ & $2.85 \times 10^{-4}$ &   2.32\\
Crab & $7.108 \times 10^{5}$ & $2.47 \times 10^{-4}$ & $3.09 \times 10^{-4}$ & $1.57 \times 10^{-4}$ & $1.23 \times 10^{-4}$ &   1.00\\\\
SNR Average & $1.077 \times 10^{6}$ & $2.22 \times 10^{-4}$ & $2.80 \times 10^{-4}$ & $1.42 \times 10^{-4}$ & $1.13 \times 10^{-4}$ &   0.92\\
\enddata
\tablenotetext{a}{Narrow line sensitivity (photons cm$^{-2}$ sec$^{-1}$)}
\tablenotetext{b}{Scale factor to apply to sensitivity plot of Fig.~\ref{crab_sens}.}
\label{sens_summ_snr}
\end{deluxetable}

\begin{deluxetable}{lccccccc}
\tabletypesize{\scriptsize}
\tablecaption{Point Source Sensitivities - Seyferts}
%\tablenum{6a}
\tablewidth{0pt}
\tablehead{\colhead{Source} & \colhead{Exposure (sec)} & \colhead{50 keV\tablenotemark{a}} & \colhead{100 keV\tablenotemark{a}}  & \colhead{511 keV\tablenotemark{a}} & \colhead{1000 keV\tablenotemark{a}} & \colhead{Scale Factor\tablenotemark{b}}}
\startdata 
1H 1934-063 & $1.070 \times 10^{6}$ & $3.16 \times 10^{-4}$ & $3.88 \times 10^{-4}$ & $1.92 \times 10^{-4}$ & $1.70 \times 10^{-4}$ &   1.38\\
3C 111 & $8.860 \times 10^{4}$ & $1.91 \times 10^{-3}$ & $2.28 \times 10^{-3}$ & $1.11 \times 10^{-3}$ & $1.02 \times 10^{-3}$ &   8.34\\
Cen A & $3.050 \times 10^{5}$ & $4.08 \times 10^{-4}$ & $5.14 \times 10^{-4}$ & $2.61 \times 10^{-4}$ & $2.13 \times 10^{-4}$ &   1.73\\
Circinus Galaxy & $8.460 \times 10^{5}$ & $2.73 \times 10^{-4}$ & $3.66 \times 10^{-4}$ & $1.85 \times 10^{-4}$ & $1.39 \times 10^{-4}$ &   1.14\\
IC 4329A & $3.050 \times 10^{5}$ & $3.48 \times 10^{-4}$ & $4.67 \times 10^{-4}$ & $2.40 \times 10^{-4}$ & $1.87 \times 10^{-4}$ &   1.53\\
IC 4518 & $1.589 \times 10^{6}$ & $3.09 \times 10^{-4}$ & $3.82 \times 10^{-4}$ & $1.89 \times 10^{-4}$ & $1.63 \times 10^{-4}$ &   1.33\\
IGR J18027-1455 & $3.938 \times 10^{6}$ & $1.97 \times 10^{-4}$ & $2.45 \times 10^{-4}$ & $1.24 \times 10^{-4}$ & $9.66 \times 10^{-5}$ &   0.79\\
MCG+08-11-011 & $9.731 \times 10^{4}$ & $1.31 \times 10^{-3}$ & $1.72 \times 10^{-3}$ & $8.67 \times 10^{-4}$ & $7.49 \times 10^{-4}$ &   6.10\\
MCG-05-23-016 & $3.452 \times 10^{5}$ & $8.75 \times 10^{-4}$ & $8.19 \times 10^{-4}$ & $3.78 \times 10^{-4}$ & $2.49 \times 10^{-4}$ &   2.03\\
MCG-06-30-015 & $3.050 \times 10^{5}$ & $3.46 \times 10^{-4}$ & $4.57 \times 10^{-4}$ & $2.34 \times 10^{-4}$ & $1.84 \times 10^{-4}$ &   1.50\\
MR2251-175 & $1.335 \times 10^{5}$ & $4.65 \times 10^{-4}$ & $6.33 \times 10^{-4}$ & $3.22 \times 10^{-4}$ & $2.83 \times 10^{-4}$ &   2.30\\
NGC 1275 & $4.132 \times 10^{4}$ & $3.18 \times 10^{-3}$ & $3.92 \times 10^{-3}$ & $1.93 \times 10^{-3}$ & $1.75 \times 10^{-3}$ &  14.27\\
NGC 4388 & $6.114 \times 10^{5}$ & $3.73 \times 10^{-4}$ & $4.47 \times 10^{-4}$ & $2.18 \times 10^{-4}$ & $1.67 \times 10^{-4}$ &   1.36\\
NGC 4507 & $9.870 \times 10^{4}$ & $6.85 \times 10^{-4}$ & $9.52 \times 10^{-4}$ & $4.90 \times 10^{-4}$ & $4.75 \times 10^{-4}$ &   3.87\\
NGC 4736 & $6.334 \times 10^{5}$ & $3.23 \times 10^{-4}$ & $4.11 \times 10^{-4}$ & $2.02 \times 10^{-4}$ & $1.78 \times 10^{-4}$ &   1.45\\
NGC 4945 & $5.932 \times 10^{5}$ & $3.87 \times 10^{-4}$ & $4.93 \times 10^{-4}$ & $2.46 \times 10^{-4}$ & $1.85 \times 10^{-4}$ &   1.51\\
NGC 5033 & $4.935 \times 10^{5}$ & $2.98 \times 10^{-4}$ & $4.13 \times 10^{-4}$ & $2.10 \times 10^{-4}$ & $1.86 \times 10^{-4}$ &   1.52\\
NGC 6300 & $1.280 \times 10^{6}$ & $3.84 \times 10^{-4}$ & $4.55 \times 10^{-4}$ & $2.33 \times 10^{-4}$ & $2.10 \times 10^{-4}$ &   1.71\\
NGC 6814 & $8.316 \times 10^{5}$ & $4.41 \times 10^{-4}$ & $5.35 \times 10^{-4}$ & $2.64 \times 10^{-4}$ & $2.30 \times 10^{-4}$ &   1.88\\\\
Seyfert Average & $1.361 \times 10^{7}$ & $7.23 \times 10^{-5}$ & $7.25 \times 10^{-5}$ & $4.23 \times 10^{-5}$ & $4.38 \times 10^{-5}$ &   0.36\\
\enddata
\tablenotetext{a}{Narrow line sensitivity (photons cm$^{-2}$ sec$^{-1}$)}
\tablenotetext{b}{Scale factor to apply to sensitivity plot of Fig.~\ref{crab_sens}.}
\label{sens_seyfert}
\end{deluxetable}

\begin{deluxetable}{lccccccc}
\tabletypesize{\scriptsize}
\tablecaption{Point Source Sensitivities - Miscellaneous/Unidentified}
%\tablenum{6a}
\tablewidth{0pt}
\tablehead{\colhead{Source} & \colhead{Exposure (sec)} & \colhead{50 keV\tablenotemark{a}} & \colhead{100 keV\tablenotemark{a}}  & \colhead{511 keV\tablenotemark{a}} & \colhead{1000 keV\tablenotemark{a}} & \colhead{Scale Factor\tablenotemark{b}}}
\startdata 
2RXP J130159 & $7.424 \times 10^{5}$ & $2.77 \times 10^{-4}$ & $3.74 \times 10^{-4}$ & $1.91 \times 10^{-4}$ & $1.41 \times 10^{-4}$ &   1.15\\
3A 1850-087 & $1.762 \times 10^{6}$ & $2.55 \times 10^{-4}$ & $3.32 \times 10^{-4}$ & $1.67 \times 10^{-4}$ & $1.47 \times 10^{-4}$ &   1.20\\
3C 273 & $3.672 \times 10^{5}$ & $2.82 \times 10^{-4}$ & $3.34 \times 10^{-4}$ & $1.95 \times 10^{-4}$ & $1.50 \times 10^{-4}$ &   1.22\\
Coma Cluster & $4.935 \times 10^{5}$ & $2.74 \times 10^{-4}$ & $3.68 \times 10^{-4}$ & $1.87 \times 10^{-4}$ & $1.66 \times 10^{-4}$ &   1.35\\
Cyg X-3 & $1.448 \times 10^{6}$ & $1.84 \times 10^{-4}$ & $2.25 \times 10^{-4}$ & $1.20 \times 10^{-4}$ & $9.99 \times 10^{-5}$ &   0.81\\
IGR J00291+5934 & $2.894 \times 10^{5}$ & $2.44 \times 10^{-3}$ & $3.08 \times 10^{-3}$ & $1.58 \times 10^{-3}$ & $1.35 \times 10^{-3}$ &  11.04\\
IGR J00370+6122 & $2.894 \times 10^{5}$ & $6.05 \times 10^{-4}$ & $8.06 \times 10^{-4}$ & $4.07 \times 10^{-4}$ & $3.49 \times 10^{-4}$ &   2.85\\
IGR J06074+2205 & $7.108 \times 10^{5}$ & $2.67 \times 10^{-4}$ & $3.42 \times 10^{-4}$ & $1.74 \times 10^{-4}$ & $1.37 \times 10^{-4}$ &   1.12\\
IGR J07597-3842 & $1.071 \times 10^{6}$ & $2.38 \times 10^{-4}$ & $2.86 \times 10^{-4}$ & $1.52 \times 10^{-4}$ & $1.20 \times 10^{-4}$ &   0.98\\
IGR J11305-6256 & $8.413 \times 10^{5}$ & $3.65 \times 10^{-4}$ & $4.66 \times 10^{-4}$ & $2.28 \times 10^{-4}$ & $1.97 \times 10^{-4}$ &   1.61\\
IGR J16167-4957 & $1.799 \times 10^{6}$ & $5.47 \times 10^{-4}$ & $6.61 \times 10^{-4}$ & $3.51 \times 10^{-4}$ & $3.23 \times 10^{-4}$ &   2.63\\
IGR J16195-4945 & $1.799 \times 10^{6}$ & $5.43 \times 10^{-4}$ & $6.98 \times 10^{-4}$ & $3.47 \times 10^{-4}$ & $3.19 \times 10^{-4}$ &   2.60\\
IGR J16479-4514 & $2.266 \times 10^{6}$ & $2.57 \times 10^{-4}$ & $3.30 \times 10^{-4}$ & $1.66 \times 10^{-4}$ & $1.52 \times 10^{-4}$ &   1.24\\
IGR J16558-5203 & $1.641 \times 10^{6}$ & $2.22 \times 10^{-4}$ & $2.85 \times 10^{-4}$ & $1.46 \times 10^{-4}$ & $1.32 \times 10^{-4}$ &   1.08\\
IGR J17252-3616 & $4.103 \times 10^{6}$ & $1.57 \times 10^{-4}$ & $2.06 \times 10^{-4}$ & $1.07 \times 10^{-4}$ & $8.51 \times 10^{-5}$ &   0.69\\
IGR J17303-0601 & $1.189 \times 10^{6}$ & $3.51 \times 10^{-4}$ & $4.49 \times 10^{-4}$ & $2.25 \times 10^{-4}$ & $1.92 \times 10^{-4}$ &   1.57\\
IGR J17597-2201 & $4.278 \times 10^{6}$ & $1.61 \times 10^{-4}$ & $2.13 \times 10^{-4}$ & $1.12 \times 10^{-4}$ & $8.66 \times 10^{-5}$ &   0.71\\
IGR J18325-0756 & $1.737 \times 10^{6}$ & $2.42 \times 10^{-4}$ & $3.10 \times 10^{-4}$ & $1.58 \times 10^{-4}$ & $1.36 \times 10^{-4}$ &   1.11\\
IGR J18450-0435 & $1.552 \times 10^{6}$ & $2.60 \times 10^{-4}$ & $3.42 \times 10^{-4}$ & $1.74 \times 10^{-4}$ & $1.55 \times 10^{-4}$ &   1.27\\
IGR J18483-0311 & $1.873 \times 10^{6}$ & $2.61 \times 10^{-4}$ & $3.38 \times 10^{-4}$ & $1.71 \times 10^{-4}$ & $1.54 \times 10^{-4}$ &   1.26\\
IGR J19140+0951 & $1.123 \times 10^{6}$ & $3.00 \times 10^{-4}$ & $3.72 \times 10^{-4}$ & $1.99 \times 10^{-4}$ & $1.84 \times 10^{-4}$ &   1.50\\
IGR J21247+5058 & $6.295 \times 10^{5}$ & $3.62 \times 10^{-4}$ & $4.63 \times 10^{-4}$ & $2.31 \times 10^{-4}$ & $1.96 \times 10^{-4}$ &   1.60\\
RX J0146 & $3.109 \times 10^{5}$ & $7.73 \times 10^{-4}$ & $9.34 \times 10^{-4}$ & $4.67 \times 10^{-4}$ & $3.91 \times 10^{-4}$ &   3.18\\
SAX J1744 & $4.270 \times 10^{6}$ & $7.38 \times 10^{-3}$ & $9.29 \times 10^{-3}$ & $4.70 \times 10^{-3}$ & $3.69 \times 10^{-3}$ &  30.09\\
SAX J1805 & $4.116 \times 10^{6}$ & $2.32 \times 10^{-4}$ & $2.90 \times 10^{-4}$ & $1.47 \times 10^{-4}$ & $1.14 \times 10^{-4}$ &   0.93\\
SS 433 & $1.261 \times 10^{6}$ & $2.38 \times 10^{-4}$ & $3.13 \times 10^{-4}$ & $1.59 \times 10^{-4}$ & $1.47 \times 10^{-4}$ &   1.20\\
V709 Cas & $2.894 \times 10^{5}$ & $2.52 \times 10^{-3}$ & $3.29 \times 10^{-3}$ & $1.63 \times 10^{-3}$ & $1.39 \times 10^{-3}$ &  11.32\\
XTE J1901+014 & $7.882 \times 10^{5}$ & $3.20 \times 10^{-4}$ & $4.20 \times 10^{-4}$ & $2.14 \times 10^{-4}$ & $1.94 \times 10^{-4}$ &   1.58\\
$\gamma$ Cas & $2.636 \times 10^{5}$ & $6.63 \times 10^{-4}$ & $8.63 \times 10^{-4}$ & $4.33 \times 10^{-4}$ & $3.60 \times 10^{-4}$ &   2.94\\
\enddata
\tablenotetext{a}{Narrow line sensitivity (photons cm$^{-2}$ sec$^{-1}$)}
\tablenotetext{b}{Scale factor to apply to sensitivity plot of Fig. ~\ref{crab_sens}.}
\label{sens_misc}
\end{deluxetable}

%%Figures*******************************
\clearpage

\epsscale{0.8}

\epsscale{1.0}


\end{document}